\documentstyle[12pt]{article}

\begin{document}
\baselineskip=20.5pt
\def\beqra{\begin{eqnarray}} \def\eeqra{\end{eqnarray}}
\def\beqast{\begin{eqnarray*}} \def\eeqast{\end{eqnarray*}}
\def\beq{\begin{equation}}      \def\eeq{\end{equation}}
\def\be{\begin{enumerate}}   \def\ee{\end{enumerate}}

\def\fnote#1#2{\begingroup\def\thefootnote{#1}\footnote{#2}\addtocounter
{footnote}{-1}\endgroup}



\def\gam{\gamma}
\def\Gam{\Gamma}
\def\la{\lambda}
\def\eps{\epsilon}
\def\La{\Lambda}
\def\si{\sigma}
\def\Si{\Sigma}
\def\al{\alpha}
\def\Th{\Theta}
\def\th{\theta}
\def\tnu{\tilde\nu}
\def\vphi{\varphi}
\def\vrho{\varrho}
\def\del{\delta}
\def\Del{\Delta}
\def\ab{\alpha\beta}
\def\om{\omega}
\def\Om{\Omega}
\def\mn{\mu\nu}
\def\mun{^{\mu}{}_{\nu}}
\def\kap{\kappa}
\def\rsi{\rho\sigma}
\def\beal{\beta\alpha}
\def\til{\tilde}
\def\rta{\rightarrow}
\def\eqv{\equiv}
\def\nab{\nabla}
\def\pa{\partial}
\def\sit{\tilde\sigma}
\def\ul{\underline}
\def\indt{\parindent2.5em}
\def\nd{\noindent}
\def\rsi{\rho\sigma}
\def\beal{\beta\alpha}
\def\caa{{\cal A}}
\def\cb{{\cal B}}
\def\cac{{\cal C}}
\def\cd{{\cal D}}
\def\ce{{\bf {\cal E}}}
\def\cf{{\cal F}}
\def\cg{{\cal G}}
\def\cah{{\cal H}}
\def\ci{{\cal I}}
\def\cj{{\cal{J}}}
\def\ck{{\cal K}}
\def\cl{{\cal L}}
\def\cm{{\cal M}}
\def\cn{{\cal N}}
\def\cO{{\cal O}}
\def\cp{{\cal P}}
\def\car{{\cal R}}
\def\cs{{\cal S}}
\def\ct{{\cal{T}}}
\def\cu{{\cal{U}}}
\def\cv{{\cal{V}}}
\def\cw{{\cal{W}}}
\def\cx{{\cal{X}}}
\def\cy{{\cal{Y}}}
\def\cz{{\cal{Z}}}
\def\asymptotic{{_{\stackrel{\displaystyle\longrightarrow}
{x\rightarrow\pm\infty}}\,\, }} 
\def\asymptext{\raisebox{.6ex}{${_{\stackrel{\displaystyle\longrightarrow}
{x\rightarrow\pm\infty}}\,\, }$}} 
\def\epsilim{{_{\textstyle{\rm lim}}\atop
_{~~~\epsilon\rightarrow 0+}\,\, }} 
\def\omegalim{{_{\textstyle{\rm lim}}\atop
_{~~~\om^2\rightarrow 0+}\,\, }} 
\def\xlimp{{_{\textstyle{\rm lim}}\atop
_{~~x\rightarrow \infty}\,\, }} 
\def\xlimm{{_{\textstyle{\rm lim}}\atop
_{~~~x\rightarrow -\infty}\,\, }} 
\def\asymptoticp{{_{\stackrel{\displaystyle\longrightarrow}
{x\rightarrow +\infty}}\,\, }} 
\def\asymptoticm{{_{\stackrel{\displaystyle\longrightarrow}
{x\rightarrow -\infty}}\,\, }} 

\def\raisenot{\raise .5mm\hbox{/}}
\def\nota{\ \hbox{{$a$}\kern-.49em\hbox{/}}}
\def\notA{\hbox{{$A$}\kern-.54em\hbox{\raisenot}}}
\def\notb{\ \hbox{{$b$}\kern-.47em\hbox{/}}}
\def\notB{\ \hbox{{$B$}\kern-.60em\hbox{\raisenot}}}
\def\notc{\ \hbox{{$c$}\kern-.45em\hbox{/}}}
\def\notd{\ \hbox{{$d$}\kern-.53em\hbox{/}}}
\def\notbd{\ \hbox{{$D$}\kern-.61em\hbox{\raisenot}}} 
\def\note{\ \hbox{{$e$}\kern-.47em\hbox{/}}}
\def\notk{\ \hbox{{$k$}\kern-.51em\hbox{/}}}
\def\notp{\ \hbox{{$p$}\kern-.43em\hbox{/}}}
\def\notq{\ \hbox{{$q$}\kern-.47em\hbox{/}}}
\def\notW{\ \hbox{{$W$}\kern-.75em\hbox{\raisenot}}}
\def\notz{\ \hbox{{$Z$}\kern-.61em\hbox{\raisenot}}}
\def\notpa{\hbox{{$\partial$}\kern-.54em\hbox{\raisenot}}}
\def\fo{\hbox{{1}\kern-.25em\hbox{l}}}  
\def\rf#1{$^{#1}$}
\def\bx{\Box}
\def\tr{{\rm Tr}}
\def\rmtr{{\rm tr}}
\def\dgg{\dagger}
\def\lag{\langle}
\def\rag{\rangle}
\def\bmid{\big|}
\def\vlap{\overrightarrow{\La p}} 
\def\lrta{\longrightarrow} \def\lrar{\raisebox{.8ex}{$\longrightarrow$}}
\def\ON{{\cal O}(N)}
\def\UN{{\cal U}(N)}
\def\bdPh{\mbox{\boldmath{$\dot{\!\Phi}$}}}
\def\bPh{\mbox{\boldmath{$\Phi$}}}
\def\bPhs{\bPh^2}
\def\sef{S_{eff}[\sigma,\pi]}
\def\sigx{\sigma(x)}
\def\pix{\pi(x)}
\def\bph{\mbox{\boldmath{$\phi$}}}
\def\bphs{\bph^2}
\def\ex{\BM{x}}
\def\exs{\ex^2}
\def\xdot{\dot{\!\ex}}
\def\y{\BM{y}}
\def\ys{\y^2}
\def\ydot{\dot{\!\y}}
\def\pat{\pa_t}
\def\pax{\pa_x}
\def\hp{{\pi\over 2}}
\def\br{{\bf r}}
\def\bk{{\bf k}}
\def\bq{{\bf q}}
\def\bx{{\bf x}}
\def\by{{\bf y}}
\def\bz{{\bf z}}
\def\bu{{\bf u}}
\def\bQ{{\bf Q}}
\def\bE{{\bf E}}
\renewcommand{\thesection}{\arabic{section}}
\renewcommand{\theequation}{\thesection.\arabic{equation}}

\begin{flushright}
\end{flushright}

\vspace*{.1in}
\begin{center}
  \Large{\sc Quantized Normal Matrices: Some Exact Results and Collective 
Field Formulation}
\normalsize

\vspace{36pt}
{\large Joshua Feinberg\fnote{*}{{\it e-mail: joshua@physics.technion.ac.il}}}
\\
\vspace{12pt}
 {\small \em Department of Physics,}\\ 
{\small \em Oranim-University of Haifa, Tivon 36006, 
Israel}\fnote{**}{permanent address}\\
{\small and}\\
{\small \em Department of Physics,}\\
{\small \em Technion - Israel Institute of Technology, Haifa 32000 Israel}

\vspace{.6cm}

\end{center}

\begin{minipage}{5.8in}
{\abstract~~~~~
We formulate and study a class of $U(N)$-invariant quantum mechanical 
models of large normal matrices with arbitrary rotation-invariant matrix
potentials. We concentrate on the $U(N)$ singlet sector of these models. In 
the particular case of quadratic matrix potential, the singlet sector can be 
mapped by a similarity transformation onto the two-dimensional 
Calogero-Marchioro-Sutherland model at specific couplings. For this 
quadratic case we were able to solve the $N-$body Schr\"odinger equation and 
obtain infinite sets of singlet eigenstates of the matrix model with given total angular 
momentum. Our main object in this paper is to study 
the singlet sector in the collective field formalism, in the large-$N$ limit. 
We obtain in this framework the ground state eigenvalue distribution and 
ground state energy for an arbitrary potential, and outline briefly the way 
to compute bona-fide quantum phase transitions in this class of models. As 
explicit examples, we analyze the models with quadratic and quartic 
potentials. In the quartic case, we also touch upon the disk-annulus quantum 
phase transition. In order to make our presentation self-contained, we also 
discuss, in a manner which is somewhat complementary to standard expositions, 
the theory of point canonical transformations in quantum mechanics for 
systems whose configuration space is endowed with non-euclidean metric, which 
is the basis for constructing the collective field theory.}

\end{minipage}

\vspace{28pt}

\vfill
\pagebreak

\setcounter{page}{1}

\section{Introduction}

Random normal matrix models were used recently in studying Laplacian growth
processes, and in particular, the fingering instability of the boundary of 
various two-dimensional fluids \cite{fingering}, the theory of the 
$\tau$-function for analytic curves \cite{tau} and two-dimensional string 
theory \cite{string}. 

An earlier important use of normal matrix models, closely related to 
the works mentioned above, was in quantum Hall physics \cite{qhe}, where it 
was shown that the partition function of the normal matrix model coincided 
with the zero-temperature partition function of two-dimensional electrons in 
strong (uniform, as well as varying) magnetic fields, and its identification 
as a complexified form of the Toda lattice $\tau$-function was established. 
(The electron quantum Hall liquid was one of the systems studied in 
\cite{fingering}.)

The structure of correlation functions in the normal matrix models was 
studied in detail in \cite{oas,chauzab} and later in \cite{wigzab}.

Normal matrices were also mentioned recently in \cite{fsz}, as a specific
kind of complex random matrices which evaded a certain generic 
geometrical constraint on the shape of the two-dimensional eigenvalue 
distribution  of a large class of circularly symmetric probability ensembles 
of complex random matrices, known as the ``Single Ring Theorem'' \cite{fz}.

All the studies mentioned above employed {\em time independent} 
normal matrices. An interesting {\em quantum mechanical} model of truly 
dynamical normal matrices, with quadratic matrix potential, was introduced in 
\cite{fs}, in their study of the physics of two-dimensional long-range Bose liquids. 
It was pointed out in \cite{fs} that the $U(N)$ singlet sector of their normal 
matrix model was equivalent to a certain two-dimensional generalization of the 
Calogero-Sutherland model \cite{calogero,sutherland}, 
which contained long-range three-body interactions, in addition to the 
repulsive two-body interactions familiar from the one-dimensional model, 
at specific values of the couplings of those interactions. (For more details 
concerning this equivalence see section 2.2.) Under this 
equivalence, the $N$ complex eigenvalues are mapped onto the positions of the 
interacting particles of the Calogero-Sutherland model in the plane, in a 
manner analogous to the relation between the Dyson matrix ensembles and the 
one-dimensional Calogero-Sutherland model \cite{sutherland}. (See also section
3 of \cite{sla}.) 

It should be noted that a three-dimensional version of the Calogero model, very
similar to the one studied in \cite{fs}, was formulated and partly solved long ago 
in \cite{cm}. Thus, we shall refer to these higher-dimensional versions of 
the Calogero model (with the Sutherland modification of the quadratic piece of the 
potential) more appropriately as Calogero-Marchioro-Sutherland  (CMS) models 
\cite{2dcs,ghosh}. We should also mention the construction of 
multidimensional Calogero models and their relation to matrix models in \cite{poly}, 
as well as the generalized multidimensional Calogero models discussed recently in 
\cite{mel}.

In this paper we shall generalize the quantum mechanical normal matrix model 
of \cite{fs} into a large class of models with $U(N)$- and rotation-invariant 
potentials. We will then analyze the singlet sector of these models in the 
large-$N$ limit, in the framework of quantum collective field theory 
\cite{jevsak, sakbook}. We will focus on studying the ground state of our generic 
quantum mechanical model. Thus, given an arbitrary $U(N)$- and rotation-invariant 
potential, we will determine the ground state eigenvalue distribution and 
ground state energy explicitly, in the large-$N$ limit. In a separate publication 
\cite{feinberg}, we will construct and analyze the effective hamiltonian of small 
fluctuations around the ground state configuration.

This paper is organized as follows. In the next section we discuss the 
quantum mechanics of the normal matrix model, concentrating on the singlet sector. 
In particular, we solve the $N-$body Schr\"odinger equation with 
quadratic matrix potential and present infinite sets of exact singlet eigenstates with 
given angular momentum. We also discuss in this section the mapping of the 
singlet sector of the normal matrix model onto the two-dimensional CMS model. 
In Section 3 we will construct the collective field quantum hamiltonian 
corresponding to the singlet sector hamiltonian $H_s$ in (\ref{singletsector}). 
The standard construction of collective field hamiltonians, as presented in 
\cite{jevsak,sakbook}, is formulated for systems defined in a flat
configuration space, endowed with cartesian coordinates. Such coordinates
are not explicitly available for our system of normal matrices, as will be explained 
in that section. In general, the transformation to the collective
hamiltonian is achieved by a quantum mechanical point canonical 
transformation from the original dynamical variables to the collective 
coordinates. Thus, we begin section 3 with a general brief discussion 
of quantum mechanical point canonical transformations in a configuration 
space endowed with a non-euclidean metric. We believe our presentation and 
results in this part of section 3 supplement those presented in chapter 
6 of \cite{sakbook}.
We then use these results to construct the collective field theory of our 
quantum mechanical model of normal matrices. We show that the Heisenberg
equations of motion (as well as the classical Hamilton equations of motion) 
of this model can be interpreted as the equations of motion of an eulerian 
fluid, similarly to an analogous interpretation of the equations of motion 
of the collective field theory of hermitean matrices \cite{matytsin}, or its
equivalent formulation as fermionic field theory \cite{polchinski}. 

Finally, in section 4 we find the ground state eigenvalue distribution 
and ground state energy of this collective hamiltonian (and thus, those of 
the original model) for an arbitrary matrix potential (\ref{potential}).
We also outline briefly the way to compute bona-fide quantum phase 
transitions in this class of models. As explicit examples, we analyze the 
models with quadratic and quartic potentials. In the quartic case, we also 
touch upon the disk-annulus quantum phase transition.

\pagebreak

\section{The Quantum Mechanical Normal Matrix Model} 
Let us first recall some basic facts about normal matrices. Consider the $N\times N$ 
complex normal matrix $M$. Thus, 
\beq\label{definition}
[M,M^\dgg] = 0\,,
\eeq
which means $M$ and $M^\dgg$ are diagonalized by the same unitary 
matrix $U$. Consequently, any normal matrix $M$ can be decomposed as 
\beq\label{decomposition}
M=U^\dgg Z U\,
\eeq
where 
\beq\label{eigenvalues}
Z = {\rm diag}~ (z_1,\ldots, z_N)
\eeq
are the complex eigenvalues. Note that $U$ in (\ref{decomposition}) is not
unique, since there is freedom to multiply it on the left by a diagonal 
unitary matrix of $N$ arbitrary phases. Thus, the count of {\em real} 
independent degrees of freedom in $M$ is the $2N$ real and imaginary 
parts of the eigenvalues, plus the $N^2$ real independent parameters of $U$, 
less the $N$ arbitrary phases, adding up to $N(N+1)$ real independent 
parameters.

The geometry of the manifold of normal matrices is defined by its embedding
in the larger space of complex matrices, with the embedding effected by the 
constraint (\ref{definition}). Thus, its line element is inherited from the 
euclidean line element $ds^2 = 
\tr\,dM^\dgg dM$ of the latter, by substituting (\ref{decomposition})
for $M$. Therefore, writing 
\beq\label{differential}
dM = U^\dgg (dZ + [dR,Z])U\,,
\eeq
where 
\beq\label{MCform}
dR = UdU^\dgg\,,\quad\quad dR^\dgg = -dR
\eeq
is the right-invariant form (i.e., invariant under $U\rightarrow UV$ with a 
fixed unitary $V$), we obtain the normal matrix line element as
\beq\label{linelement}
ds^2 = \sum_i |dz_i|^2 + 2\sum_{i<j} dR_{ij}^*\,dR_{ij} \,|z_i - z_j|^2\,.
\eeq
Note that this line element is independent of the diagonal elements $dR_{ii}$ 
(which are essentially the differentials of the $N$ arbitrary phases mentioned
above). It depends only on the differentials of the real and imaginary parts of the 
$z_i$, and on the differentials of the real and imaginary parts of the 
elements in the lower-triangular part 
of $dR$, $N(N+1)$ in all. Note also that there are no metric elements in 
(\ref{linelement}) which mix $dz$'s and $dR$'s: motions tangential to 
eigenvalues and to the unitary group are orthogonal.

The laplacian associated with the metric (\ref{linelement}) is thus the sum 
of two terms 
\beq\label{nablasquare}
\nabla^2 = \nabla_s^2 + \nabla_{U(N)}^2\,,
\eeq
where $\nabla_s^2$ 
arises from pure eigenvalue variations and $\nabla_{U(N)}^2$ arises from pure 
$U(N)$ rotations. In particular, $\nabla_{U(N)}^2$ annihilates $U(N)$ 
singlets. Thus, $\nabla^2 = \nabla_s^2$ in the $U(N)$ singlet sector. 
Clearly, $\nabla_s^2$ and $\nabla_{U(N)}^2$ are analogous, respectively, to 
the radial and to the ${\bf L}^2/r^2$ parts of the laplacian in atomic 
physics.

The explicit expressions for $\nabla_s^2$ and $\nabla_{U(N)}^2$ can be 
computed from (\ref{linelement}) in a straightforward manner. The angular part
of the laplacian is 
\beq\label{angularpart}
\nabla_{U(N)}^2 = 2\sum_{i<j} {{\partial\over \partial R_{ij}}
{\partial\over \partial R_{ij}^*}\over |z_i-z_j|^2}\,,
\eeq
and the singlet part is 
\beq\label{singletpart}
\nabla_s^2 = {1\over |\Delta|^2}\sum_i\left[{\partial\over\partial x_i}
\left( |\Delta|^2 {\partial\over\partial x_i}\right) + 
{\partial\over\partial y_i}\left( |\Delta|^2 {\partial\over\partial y_i}
\right)\right]\,,
\eeq
where
\beq\label{vandermonde}
\Delta = \prod_{i>j} (z_i - z_j) 
\eeq
is the Vandermonde determinant of eigenvalues, and $z_i = x_i + iy_i$.   
In terms of the complex derivatives 
\beq\label{complexderivative}
\pa_i = {\partial\over\partial z_i} = \frac{1}{2} 
({\partial\over\partial x_i} - i{\partial\over\partial y_i})\,,\quad\quad
\pa_i^* = {\partial\over\partial z_i^*} = \frac{1}{2} 
({\partial\over\partial x_i} + i{\partial\over\partial y_i})
\eeq
we can write $\nabla_s^2$ more elegantly as  
\beq\label{singletpartholomorphic}
\nabla_s^2 = 2 \sum_i\left( {1\over \Delta^*} \pa_i^* \Delta^* \pa_i 
 + {1\over \Delta} \pa_i \Delta \pa_i^*\right)\,.
\eeq

In this paper we study $U(N)$- and rotation-invariant normal matrix models defined 
by quantum hamiltonians of the general form 
\beq\label{hamiltonian}
H = -{1\over 2}\nabla^2 + \tr V(M^\dgg M)\,.
\eeq
Here $V$ is a generic polynomial potential, whose couplings scale properly 
with $N$, in order to ensure a well-behaved large-$N$ limit. We will see
later-on that the required $N$-dependence is 
\beq\label{potential}
V(M^\dgg M) = \sum_{p\geq 1} {g_p(M^\dgg M)^p\over N^{p-1}}\,,
\eeq
(with $N$-independent $g_p$).

According to general principles, the ground state of (\ref{hamiltonian}) 
must be a $U(N)$ singlet. As was stated earlier, we will find this ground state 
explicitly in the large-$N$ limit. Thus, in this paper we shall focus exclusively on the 
$U(N)$-singlet sector of (\ref{hamiltonian}), in which this hamiltonian 
is reduced to
\beq\label{singletsector}
H_s = -{1\over 2}\nabla_s^2 + \sum_i V(|z_i|^2)\,,
\eeq
which defines the dynamics of the $N$ complex eigenvalues $z_i$. It acts on 
singlet wave functions $\chi_s(z_1,z_1^*, \ldots , z_N, z_N^*)$ which 
are completely symmetric under eigenvalue permutations.

Note from (\ref{singletpartholomorphic}) that $H_s$ is inviariant under the interchange 
$z_i, \pa_i \leftrightarrow z_i^*, \pa_i^*$, i.e., under reflection with respect to the $x$-axis.
This means that if $\chi_s$ is an eigenstate of $H_s$ with eigenvalue $E$, so is $\chi_s^*$,
\beq\label{degeneracy} 
H_s\chi_s = E\chi_s \iff H_s\chi_s^* = E\chi_s^*\,.
\eeq
Thus, if $\chi_s$ is complex (and not just a real wave function multiplied by a phase), its 
corresponding energy will be at least doubly degenerate.

It is clear from the metric (\ref{linelement}) that $H_s$ is symmetric 
$\langle \chi_{s1}|H_s\chi_{s2}\rangle = \langle H_s\chi_{s1}|\chi_{s2}\rangle $ with respect 
to the measure 
\beq\label{complexmeasure}
d\mu\left(\{z_i,z_i^*\}\right) = |\Delta|^2 \prod_{k=1}^N d^2 z_k
\eeq
and the inner product 
\beq\label{singletinnerprod}
\langle \chi_{s1}|\chi_{s2}\rangle = \int\,d\mu\,\chi_{s1}^*\, \chi_{s2}\,. 
\eeq
As is evident from (\ref{singletpartholomorphic}) (or (\ref{singletpart})), all
interparticle interactions are lumped into the kinetic term of $H_s$, 
through kind of a complex gauge field 
\beq\label{logarithmicderivative}
A_i\left(\{z_k\}\right) = 
{\pa_i\Delta\over \Delta} = \sum_j\!'{1\over z_i-z_j}
\eeq
(where the prime indicates that the sum is over all $j\neq i$), and
its complex conjugate field $A_i^* = A_i\left(\{z_k^*\}\right)$, which multiply
the first order derivatives $\pa_i^*$ and $\pa_i$, respectively,
\beq\label{singletpartholomorphic1}
\nabla_s^2 = 2 \sum_i\left(2\pa_i^*\pa_i + A_i^*\pa_i + A_i\pa_i^* \right)\,.
\eeq

\subsection{Exact Eigenstates of the Matrix Model With Quadratic Potential}
Let us consider now (\ref{singletsector}) with quadratic potential
\beq\label{quadraticpotential1}
\tr V(M^\dgg M) = {1\over 2}m^2\, \tr M^\dgg M = {1\over 2}m^2 \,\sum_i |z_i|^2 
\,,\quad\quad m^2 > 0\,.
\eeq
Remarkably, infinite (albeit incomplete) sets of eigenstates of this singlet 
hamiltonian $H_s$ can be found analytically, in a similar manner\footnote{In contrast
with (\ref{quadraticpotential1}), the quadratic potential term in \cite{cm} was 
$\propto \sum_{i<j} |z_i - z_j|^2$.} to \cite{cm}. 
\subsubsection{Radial Eigenstates}
It is possible to compute explicitly the ground state $\chi_0$ of this matrix model, 
and also an infinite set of radial excitations above it. To this end, define the 
radial combination  
\beq\label{vrho}
\vrho = \sum_{i=1}^N |z_i|^2\,, 
\eeq
which is symmetric under permutation of eigenvalues, and look for eigenstates of $H_s$ 
of the form $\chi = F(\vrho)$. Using the identity 
\beq\label{logidentity}
\sum_{i=1}^N\,z_i\,\pa_i\log \Delta = {N(N-1)\over 2}
\eeq
and a similar identity for $\Delta^*$, we can write the Schr\"odinger equation 
$H_s F(\vrho) = E F(\vrho) $  with potential (\ref{quadraticpotential1}) as 
\beq\label{Fschrodingereq}
2\vrho\,{d^2F\over d\vrho^2} + N(N+1)\,{dF\over d\vrho} + \left(E -{1\over 2}m^2\,
\vrho \right) F = 0\,.
\eeq
The eigenfunction solutions of (\ref{Fschrodingereq}) and their corresponding 
eigenvalues are readily found as 
\beq\label{eigensolutions}
\chi_n (\vrho) = L_n^{({N(N+1)\over 2} -1)} (m\vrho) \, 
e^{-{1\over 2}m\vrho}\,,\quad\quad E_n = \left({N(N+1)\over 2} + 2n \right)m\,,
\eeq
$(n= 0,1,\ldots )$, where 
\beq\label{laguerre}
L_n^{({N(N+1)\over 2} -1)} (u) = \sum_{k=0}^n\,{(-1)^k\over k!}\,
{\left({N(N+1)\over 2} + n -1\right)! \over (n-k)!\,
\left({N(N+1)\over 2} + k -1\right)!}\,u^k
\eeq
is a Laguerre polynomial. 

In particular, the nodeless gaussian eigenfunction $\chi_0 = e^{-{1\over 2}m\vrho} = 
e^{-{1\over 2}m \sum_i |z_i|^2 } $ is the ground state \cite{2dcs}, with energy 
\beq\label{exactgsenergy}
E_0 = {N(N+1)\over 2}\,m\,.
\eeq
The radial states of higher energy are evenly spaced, with constant gap $2m$.

From (\ref{complexmeasure}) and (\ref{eigensolutions}), we obtain that the joint 
probability density of the $N$ eigenvalues in the ground state is proportional to 
\beq\label{gsjpd}
|\Delta \chi_0|^2 = \Big|\prod_{i>j} (z_i - z_j)\Big|^2 e^{-m \sum_i |z_i|^2 }\,.
\eeq
Starting from the two-dimensional CMS model (at couplings tuned to the point 
corresponding to the normal matrix model), it was shown in \cite{2dcs} 
(prior to \cite{fs}) that the ground state wave function of the CMS model was 
simply 
\beq\label{CMSgs}
\psi^{CMS}_0 = \Big|\prod_{i>j} (z_i - z_j)\Big| \chi_0\,,
\eeq
which thus led to a joint probability density for the positions of the $N$ particles,
which was also proportional to (\ref{gsjpd}). It was further pointed out in 
\cite{2dcs} that (\ref{gsjpd}) 
was also proportional to the joint probability density of the $N$ eigenvalues in 
Ginibre's ensemble \cite{ginibre} of gaussian random complex matrices, which also 
happens to coincide with the joint probability density of the gaussian 
(time independent) normal matrix model. This is, of course, in complete analogy  
with the one-dimensional case, in which the joint probability density for 
the positions of the $N$ particles in the ground state of the Calogero-Sutherland 
model at specific values of the repulsive interaction, coincides with that of the 
appropriate Dyson random matrix ensemble \cite{sutherland}.

It was noted in \cite{ghosh}\footnote{This was done in section 2 of \cite{ghosh}, in the 
context of the two-dimensional CMS model, and not directly in the normal matrix model. 
The adaptation of the discussion in \cite{ghosh} to the case of our normal matrix model
is what follows.} that the 
eigenstates (\ref{eigensolutions}) could be also obtained using group theoretical 
considerations. As it turns out, the normal matrix model with quadratic 
potential\footnote{In the following group theoretical discussion we will set $m=1$ in 
(\ref{quadraticpotential1}) for convenience (but will reinstate it back following 
(\ref{chin})).} (\ref{quadraticpotential1}) has dynamical 
symmetry $O(2,1)$
\beq\label{o21}
[h,D] = ih\,,\quad [h,K] = 2iD\,,\quad {\rm and}\quad [K,D] = -iK\,,
\eeq
generated by 
\beqra\label{o21generators}
h &=& -{1\over 2} \nabla_s^2 = - \sum_i\left( {1\over \Delta^*} \pa_i^* \Delta^* \pa_i 
 + {1\over \Delta} \pa_i \Delta \pa_i^*\right)\nonumber\\{}\nonumber\\
K &=& {1\over 2} \vrho = {1\over 2}\sum_{i=1}^N |z_i|^2 \nonumber\\{}\nonumber\\   
D &=& {i\over 4} \sum_i\left( \{z_i,\pa_i\} + \{z_i^*,\pa_i^*\}  + N - 1\right)\,.
\eeqra
The operator $h$ is just the kinetic part of $H_s$, $K$ is the conformal generator, and
$D$ generates dilatations. Let us define the raising and lowering operators $B^\pm$ as
\beq\label{rasinglowering}
B^\pm = {1\over 2} \left( h -K \mp 2iD\right)\,.
\eeq
These operators, together with the full hamiltonian $H_s = h+K$ generate the $SU(1,1)$ 
algebra
\beq\label{su11}
[H_s,B^\pm] = \pm 2B^\pm\,,\quad [B^-,B^+] = H_s\,.
\eeq
Then, the states (\ref{eigensolutions}) are obtained by applying $B^+$ repeatedly to 
the ground state $\chi_0$, namely, 
\beq\label{chin}
\chi_n = (B^+)^n\,\chi_0\,,
\eeq
from which it is easy to check that indeed $E_n = E_0 + 2mn\,.$

\subsubsection{Eigenstates with Nonvanishing Angular Momentum}
The eigenstates (\ref{eigensolutions}) are rotation invariant and thus do not carry 
angular momentum. Here we derive an infinite set of eigenstates which do carry 
angular momentum. The derivation of these states is similar to the derivation of the 
radial states. 

We shall look here for eigenstates of the form $F(\vrho)\,g(z_1,\ldots z_N)$,
with $g(z_1,\ldots z_N)$ a totally symmetric holomorphic function. Using the identity 
(\ref{logidentity}) (and the corresponding one for $\Delta^*$), we 
can write the Schr\"odinger equation as 
\beq\label{Fgschrodingereq1}
\left[2\vrho\,{d^2F\over d\vrho^2} + N(N+1)\,{dF\over d\vrho} + \left(E -{1\over 2}m^2\,
\vrho \right) F\right]\,g + 2\,{dF\over d\vrho}\,\sum_i (z_i\pa_i g) + F\,\sum_i 
\left(\pa_ig\,{\pa_i^* \Delta^*\over \Delta^*} \right) = 0\,.
\eeq
From the identity $\sum_i {\pa_i^* \Delta^*\over \Delta^*} =0$ we see that if we choose 
the symmetric function $g$ such that $\pa_i g$ will be independent of the index $i$, and 
furthermore, if we choose it to be homogeneous of some degree $p$ such that 
$\sum_i (z_i\pa_i g) = pg$, then (\ref{Fgschrodingereq1}) will imply that 
\beq\label{Fgschrodingereq}
2\vrho\,{d^2F\over d\vrho^2} + \left(N(N+1)+2p\right)\,{dF\over d\vrho} + 
\left(E -{1\over 2}m^2\,\vrho \right) F = 0\,,
\eeq
which coincides with (\ref{Fschrodingereq}) up to shifting $N(N+1)$ by $2p$. 
The two restrictions on $g$ uniquely fix it (up to a multiplicative constant) as 
$g(z_1,\ldots z_N) = (\sum_i z_i)^p$. Thus, by comparing (\ref{Fgschrodingereq}) and 
(\ref{Fschrodingereq}) we see that 
\beq\label{Fgeigensolutions}
\chi_n^{(p)} = (z_1 + \ldots + z_N)^p\,L_n^{({N(N+1)\over 2} +p-1)} 
(m\vrho) \, e^{-{1\over 2}m\vrho}
\eeq
is an eigenfunction of $H_s$, with corresponding eigenvalue 
\beq\label{Fgeigenvalues}
E_n^{(p)} = \left({N(N+1)\over 2} + p + 2n \right)m
\eeq
$(n= 0,1,\ldots )$. The radial eigenstates (\ref{eigensolutions}) are just the states 
(\ref{Fgeigensolutions}) with $p=0$. Of course, $\chi_n^{(p)*}$ is also an eigenstate, 
linearly independent of $\chi_n^{(p)}$ (for $p\neq 0$), and with the same eigenvalue 
$E_n^{(p)}$, in accordance with (\ref{degeneracy}). 

By applying the angular momentum operator 
\beq\label{angularmomentumoperator}
L_z = \sum_i(z_i\pa_i - z_i^*\pa_i^*)
\eeq
to $\chi_n^{(p)}$ we see that it carries angular momentum $p$. Thus, $p$ is a nonegative 
integer. Similarly, $\chi_n^{(p)*}$ carries angular momentum $-p$. It is reasonable to 
expect that analogous states should exist also in the two-dimensional CMS model at 
arbitrary couplings.

The generalized multi-dimensional Calogero model discussed recently in \cite{mel} allows
for $N$ distinguishable particles in arbitrary number of dimensions. Part of the spectrum 
of that model was studied using group theoretical methods. In two dimensions, and for 
the special case of identical particles, the model in \cite{mel} can be mapped by a 
similarity transformation onto our singlet sector hamiltonian (\ref{singletsector}) 
with quadratic potential (\ref{quadraticpotential1}). (See the discussion following 
(\ref{similarH}).) Our eigenstates (\ref{Fgeigensolutions}) coincide with a certain 
subset of the eigenstates obtained in \cite{mel} upon tuning the couplings of the latter 
model to coincide with ours.

By comparing (\ref{Fgeigenvalues}) and (\ref{eigensolutions}), we see that the pair of 
first excited states among the states we have constructed is the one corresponding to 
$n=0$ and $p=1$, with energy difference  
\beq\label{energygap}
E_0^{(1)} - E_0^{(0)} = m
\eeq
above the ground state, which thus sets an {\em upper bound} on the gap of excitations 
in the spectrum of $H_s$. This upper bound on the gap, obtained from {\em exact} 
singlet eigenstates, is half the gap computed recently in \cite{2dcscollective} in the 
large-$N$ limit, using collective field analysis, for the equivalent CMS model at 
the normal matrix point in coupling space. 

\subsubsection{The Quantum Normal Matrix Model and Physics of Electrons in the Lowest Landau Level}
The joint probability density (\ref{gsjpd}) for $N$ eigenvalues in the ground state 
$\chi_0$ coincides, upon identification of ${1\over\sqrt{2m}}$ as the magnetic length,  
with that of $N$ noninteracting electrons which occupy the lowest Landau Level (LLL) 
(at minimal angular momentum). Indeed, $\Delta \chi_0$ can be identified with the 
$\nu=1$ Laughlin wave function.

However, this relation between the ground state of $H_s$ with quadratic potential and 
many-body wave functions in the LLL, does not persist to the excited 
states $\chi_n\, (n\geq 1)$ and $\chi_n^{(p)*}\, (p\geq 1)$ in (\ref{eigensolutions}) 
and (\ref{Fgeigensolutions}), for a very simple reason. Our quantum mechanical normal matrix 
model is invariant under reflection with respect to the $x$-axis, which leads to the 
degeneracy (\ref{degeneracy}). In contrast, in the quantum Hall problem, $N$-electron 
states in the LLL (in the symmetric gauge) are in the form of an antisymmetric holomorphic 
function $f_A(z_1,\ldots,z_N)$ ( multiplied by the gaussian $e^{-{1\over 2}m\vrho}$). Any 
dependence in $f_A$ on nonhomolorphic variables $z_i^*$ means that the corresponding 
state belongs to higher Landau levels\footnote{As is well-known, this happens because the Landau 
hamiltonian in the symmetric gauge is the sum of a rotation invariant piece and a piece 
proportional to $BL_z  = B\sum_i(z_i \pa_i - z_i^*\pa_i^*)$, which breaks the symmetry under
interchanging $z_i,\pa_i\leftrightarrow z_i^*, \pa_i^*$.}. Thus, for example, antiholomorphic 
wave functions will correspond to higher energies than their holomorphic counterparts, unlike
(\ref{degeneracy}). As a particular example, consider the pairs of degenerate eigenstate  
$\chi_0^{(p)}, \chi_0^{(p)*} $ of $H_s$. The wave function  $\Delta\,\chi_0^{(p)}$ is of course 
a linear combination of LLL states\footnote{The $p=1$ case is special, since 
$\Delta\,\chi_0^{(1)}$ can be written as a single determinant (rather than a linear combination
of determinants) of holomorphic monomials $z^n$, multiplied by gaussian factors, of the form 
$\tilde \Delta \chi_0$, where $\tilde \Delta$ differs from the Vandermonde determinant by 
having its last row entries $z_i^N$ instead $z_i^{N-1}$.}, but $\Delta\,\chi_0^{(p)*}$ is of 
higher energy in the Landau problem.

In very intense magnetic fields, which in our normal matrix model corresponds to large $m^2$, 
the gap from the LLL to the next Landau level becomes very large, and low energy physics can be 
described entirely in terms of antisymmetric holomorphic wave functions. 
This means that the number of degrees of freedom is cut in half: to each electron corresponds 
the combination $z_i = x_i + iy_i$ of position operators, and not $x_i$ and $y_i$ separately. 
However, in our inverstigation of the singlet sector of the normal matrix model in this paper we 
do not have any special interest in taking the limit $m\rightarrow\infty$ so as to make the gap 
to the excited states infinite. Thus, we will have to take into account all singlet states, 
holomorphic as well as nonholomorphic.

In fact, it might be the case that a version of the quantum normal matrix model considered in 
this paper, in which a random piece is added to the potential $V(|z|^2)$, could teach us 
something relevant for understanding the levitation (or floating) problem of the extended states 
away from the center of the Landau band in quantum Hall physics, at weak magnetic fields.

In order to relate our quantum normal matrix model and the time-independent normal matrix
models \cite{fingering,qhe}, however, the large $m^2$ limit is important, precisely because 
of the large energy gap, proportional to $m$, separating the ground state $\chi_0$ and the 
excited states of $H_s$. Thus, at large $m^2$, the low-energy physics of $H_s$ 
(with $V(|z|^2) = {1\over 2}m^2 |z|^2$) is captured entirely by the ground state. As was 
mentioned earlier, according to \cite{qhe}, the 
joint probability density of $N$ noninteracting electrons, which occupy the LLL in a 
strong (and possibly nonuniform) magnetic field, coincides with the joint probability 
density of the eigenvalues of the zero-dimensional normal matrix model, which is 
given by 
\beq\label{chaunormal}
P(z_1,\ldots , z_N)  = {1\over Z}\,\Big|\prod_{i>j} (z_i - z_j)\Big|^2
\,e^{-\sum_i W(|z_i|^2)}\,,
\eeq
where $W(|z|^2)$ is the magnetostatic potential, namely, $B(z,z^*)\propto \pa\pa^* W(|z|^2)$. 
This result was obtained by observing that the $N$-electron wave function, with all spins 
parallel and with minimal angular momentum (${N(N-1)\over 2})$, was simply 
\beq\label{LLL}
\psi_{LLL} = \prod_{i>j} (z_i - z_j) \,e^{-{1\over 2}\sum_i W(|z_i|^2)}\,,
\eeq
i.e., a generalized form of the $\nu=1$ Laughlin wave function. For a gaussian 
$W ={1\over 2}m^2 |z|^2 $, which corresponds to a uniform magnetic field, $\psi_{LLL}$ is 
just the function $\Delta\chi_0$, as we saw above.

Note, however, that for a polynomial $W(|z|^2)$ of higher degree,
${\psi_{LLL}\over \Delta} = e^{-{1\over 2}\sum_i W(|z_i|^2)}$ is not an 
eigenstate of $H_s$ (\ref{singletsector}) with a {\em local} potential $V(|z|^2)$. 
Thus, in nonuniform magnetic fields, we cannot relate $\psi_{LLL}$ to the 
ground state of $H_s$ with any reasonable potential. 

As the final comment in this section, note that the {\em many-body} wave function $\chi_n^{(p)}$ 
is somewhat reminescent of the {\em single} particle wave function in the $n$th Landau 
level and with angular momentum $p$, which is given (in the symmetric gauge) by 
$z^p L_n^{(p)} (m\vrho) e^{-{1\over 2}m\vrho}$ (where in the latter expression 
$\vrho = |z|^2$).

\subsection{Relation to the Two-Dimensional Calogero-Marchioro-Sutherland Model}
The hamiltonian of the CMS model is \cite{cm,2dcs}
\beq\label{hcms}
H_{CMS} = -{1\over 2} \sum_i \nabla_i^2 + g\sum_{i<j} {1\over \br_{ij}^2} + G\sum_i
\sum_{j<k}\!' {\br_{ij}\cdot\br_{ik} 
\over \br_{ij}^2 \br_{ik}^2} + {m^2\over 2} \sum_i \br_i^2
\eeq
where $\br_{ij} = \br_i - \br_j$. Thus, $H_{CMS}$ contains a three-body long-range 
interaction, in addition to the repulsive two-body long-range interaction familiar from 
the one-dimensional Calogero-Sutherland model.

In order to see the equivalence between our normal matrix model and the two-dimensional
CMS model (at the specific point in coupling space) we have to remove the 
``gauge fields'' $A_i$ and $A_i^*$ (\ref{logarithmicderivative}) from 
(\ref{singletpartholomorphic1}). These fields can be gauged away by a nonunitary gauge 
transformation, or more precisely, by a (singular) similarity transformation. 
Indeed, it is a straightforward calculation to show that 
\beq\label{singletsimilarity}
\nabla_s^2 = {1\over |\Delta|}\left(4\sum_i \pa_i^*\pa_i - \sum_i 
{\pa_i\Delta\over \Delta}{\pa_i^*\Delta^*\over \Delta^*}\right)|\Delta|\,.
\eeq
Thus, 
\beq\label{singletHsimilarity}
H_s = {1\over |\Delta|}\left(
-{1\over 2}\sum_i \nabla_i^2 + \sum_i V(|z_i|^2)
+{1\over 2}\sum_i 
{\pa_i\Delta\over \Delta}{\pa_i^*\Delta^*\over \Delta^*}
\right)|\Delta|\,,
\eeq
where we used $4\pa_i^*\pa_i = {\pa^2\over \pa x_i^2} + 
{\pa^2\over \pa y_i^2} = \nabla_i^2$. 
The hamiltonian obtained from $H_s$ after the similarity transformation, 
\beq\label{similarH}
{\cal H}_s = |\Delta| H_s {1\over |\Delta|} = 
-{1\over 2}\sum_i \nabla_i^2 + \sum_i V(|z_i|^2)
+{1\over 2}\sum_i 
{\pa_i\Delta\over \Delta}{\pa_i^*\Delta^*\over \Delta^*}\,,
\eeq
has the standard kinetic term. Thus, we have removed the gauge fields in 
(\ref{singletpartholomorphic1}) at the price of introducing the interaction terms 
$\sum_i {\pa_i\Delta\over \Delta}{\pa_i^*\Delta^*\over \Delta^*}$.

In order to proceed we note, using (\ref{logarithmicderivative}), that 
\beqra\label{observation}
\sum_i {\pa_i\Delta\over \Delta}{\pa_i^*\Delta^*\over \Delta^*} &=& 
\sum_i\sum_j\!'{1\over z_i-z_j}\sum_k\!'{1\over z_i^*-z_k^*} 
\nonumber\\
&=& 2\sum_{i<j} {1\over |z_i-z_j|^2} + 2\sum_i
\sum_{j<k}\!' {{\rm Re} \left((z_i-z_j)^*(z_i-z_k)\right)
\over |z_i-z_j|^2 |z_i-z_k|^2}\nonumber\\
&=& 2\sum_{i<j} {1\over \br_{ij}^2} + 2\sum_i
\sum_{j<k}\!' {\br_{ij}\cdot\br_{ik} 
\over \br_{ij}^2 \br_{ik}^2}\,,
\eeqra
where we denoted the position operator of the $i$th eigenvalue $z_i=x_i + iy_i$ 
in the plane by $\br_i = x_i\hat \bx + y_i\hat \by\,.$
Substituting (\ref{observation}) in (\ref{similarH}) and comparing it with (\ref{hcms}),
we see that for the particular potential $V(|z|^2) = {1\over 2} m^2 |z|^2$, 
the hamiltonian ${\cal H}_s$ coincides with $H_{CMS}$ at $g=G=1$ \cite{fs,2dcs,ghosh}.

The CMS model (\ref{hcms}), at the special point $g=G=1$, was studied recently in 
\cite{2dcscollective}, in the collective field approach in the large-$N$ 
limit, and it was found that the particles condensed into a uniform 
disk, in accordance with the findings of \cite{2dcs}. The authors of 
\cite{2dcscollective} went one step farther, and also computed fluctuations around the 
uniform ground state configuration, which required taking the first subleading term in 
the large-$N$ expansion. They have found that the spectrum of those fluctuations had 
an energy gap equal to $m$, corresponding to setting $V(|z|^2) = {1\over 2} m^2 |z|^2$ 
in (\ref{similarH}), as was already briefly mentioned following (\ref{energygap}). 
In this work we extend the leading large-$N$ results of \cite{2dcscollective} to the 
case of an arbitrary external potential.

The hamiltonian $H_s$ acts on singlet wave-functions 
$\chi_s$, which are completely symmetric under permutations of the $z_i$. The 
similarity transformation on $H_s$ in (\ref{similarH}) implies a similarity 
transformation $\chi_s\rightarrow |\Delta|\chi_s(z_1,z_1^*\ldots, z_N,z_N^*)$ into wave 
functions which are also completely symmetric under permutations of the eigenvalues.
Thus, ${\cal H}_s$ in (\ref{similarH}), with its conventional flat space kinetic term, 
corresponds to a system of interacting bosons in two dimensions. In contrast, in the corresponding 
quantum mechanical hermitean matrix models, the analog of ${\cal H}_s$ acts on one-dimensional 
non-interacting fermions, as was shown in \cite{bipz}. 

Note, however, that the similarity trasformation (\ref{similarH}), which maps the matrix model 
hamiltonian $H_s$ onto the CMS hamiltonian ${\cal H}_s$ is highly singular, since $|\Delta| = 
\Big|\prod_{i>j} (z_i - z_j)\Big|$ has branch points wherever two eigenvalues coincide. 
This might lead to some delicate issues in trying to decide whether the singlet sector 
of the normal matrix model and the CMS model are completely equivalent or not. 
There is no such problem in the mapping of the singlet sector of the quantum mechanical 
hermitean matrix models onto one-dimensional noninteracting fermions. 

We stress that our construction of the quantum collective field formulation of the 
singlet sector of the normal matrix model in section 2.2 is based entirely on 
the non-euclidean configuration space with metric (\ref{linelement}) and the corresponding 
hamiltonian $H_s$. Nowhere in our construction do we resort to the singular similarity
transformation into euclidean configuration space and CMS hamiltonian ${\cal H}_s$. 
Nevertheless, as we have already remarked above, the results of our collective field 
analysis (for the particular potential $V(|z|^2) = {1\over 2} m^2 |z|^2$) agree 
(to leading order in ${1\over N}$) with those of \cite{2dcscollective}, which were 
obtained starting with ${\cal H}_s$.

\subsection{A Concluding Remark Concerning Cartesian Coordinates and Feynman Diagrams}
We end this section with the following comment: 
It is customary to refer to (\ref{decomposition}) as the {\em polar decomposition} of 
$M$. Thus, the $z_i$ and $R_{ij}$ ($i<j$) are referred to as the polar coordinates
of $M$. The matrix elements $M_{ij}$ are known as the cartesian coordinates. 
Thus, (\ref{linelement}), and (\ref{angularpart}), (\ref{singletpart}) express
the metric and the laplacian in polar coordinates. There are no simple 
expressions for these objects of normal matrix theory in terms of 
the cartesian coordinates $M_{ij}$, since it is rather complicated
to solve the constraint (\ref{definition}) in these coordinates explicitly. 
To this end one has to work in the bigger space of complex matrices and 
introduce a matrix Lagrange multiplier field to impose (\ref{definition}) on 
the dynamics, and then follow Dirac's procedure for constrained quantization.
Interestingly enough, these complications do not arise for hermitean matrices,
which is of course a subclass of normal matrices. In this case, the 
{\em linear} constraint $M_{ij} = M_{ji}^*$ implies simply 
$\nabla^2 = \sum_{ij} \partial^2/(\partial M_{ij}\partial M_{ij}^*) = 
\tr (\partial^2/\partial M^2)$. 

Having a lagrangian formulation of the normal matrix model in cartesian 
coordinates is necessary in order to formulate the large-$N$ ``double-line'' diagrammatic 
expansion of the model. In this respect, it should be noted that this expansion will be in 
powers of ${1\over N}$ rather than powers of  ${1\over N^2}$, as should be clear from the 
exact expression (\ref{exactgsenergy}) for the ground state energy in the case of a 
quadratic potential. 

\pagebreak

\section{Construction of the Collective Hamiltonian}
\setcounter{equation}{0}
The transformation from the hamiltonian in its original form
to its expression in terms of collective coordinates is achieved by a 
quantum mechanical point canonical transformation from the original dynamical
variables to collective coordinates. 

The configuration space of our quantum mechanical normal matrix model 
is parametrized by the polar coordinates $z_i$ and $R_{ij} (i<j)$,
and is endowed with the non-euclidean metric (\ref{linelement}). 

Thus, in order to prepare the ground for the construction of the 
collective field hamiltonian for the normal matrix model, we will first
recall some basic facts about point canonical transformations in 
configuration spaces whose geometry is defined by a non-euclidean metric,
thus extending the discussion in chapter 6 of \cite{sakbook}. 

\subsection{Point Canonical Transformations in Non-Euclidean Configuration 
Space}

We shall start by stating some elementary facts, mainly in order to introduce
notations. We make our points starting in Eq.(\ref{Heff}). 

Consider a quantum mechanical system, defined in a D-dimensional 
configuration space ${\cal C}$, with coordinates $q^a (a=1,\ldots D)$ 
and line element $ds^2 = g_{ab}(\bq)dq^a dq^b$. The hamiltonian is 
given by\footnote{In this subsection we will write all factors of $\hbar$ 
explicitly, in order to identify the purely quantum mechanical terms induced 
by the transformation in the effective hamiltonian, defined later in 
this section.}
\beq\label{hq}
H_q = -{\hbar^2\over 2}\nabla_q^2 + V(\bq)\,,
\eeq
where
\beq\label{qlaplacian}
\nabla_q^2 = {1\over\sqrt{g}}{\pa\over\pa q^a}\left(g^{ab}\sqrt{g}
{\pa\over\pa q^b}\right) 
\eeq
is the invariant laplacian, expressed in terms of the coordinates $q^a$. 
In (\ref{qlaplacian}) $g^{ab}$ is the inverse metric and $g=\det g_{ab}$. This laplacian (and the corresponding hamiltonian
$H_q$) are symmetric $\langle \psi_1|H_q\psi_2\rangle =
\langle H_q\psi_1|\psi_2\rangle $ with respect to the measure 
$\sqrt{g}\,d\bq$ and the inner product $\langle \psi_1|\psi_2\rangle 
= \int\limits_{\cal C} \sqrt{g}\,d\bq \,\psi_1^*(\bq)\,\psi_2(\bq)$.

Let us consider a point canonical transformation 
\beq\label{qQtransformation}
q^a\rightarrow Q^a = Q^a (\bq)\,,
\eeq
and assume its inverse exists as well: 
\beq\label{Qqtransformation}
q^a = q^a(\bQ)\,.
\eeq
We know from rudimentary differential geometry, that since the line element 
$ds^2$ is invariant under this transformation, the metric in the new 
coordinates is given by
\beq\label{Qmetric}
\Om_{ab} ( \bQ) = g_{mn}(\bq(\bQ)){\partial q^m\over\partial Q^a}
{\partial q^n\over\partial Q^b}\,,
\eeq
its inverse by 
\beq\label{inverseQmetric}
\Om^{ab} ( \bQ)=  g^{mn}(\bq(\bQ)){\partial Q^a \over \partial q^m}
{\partial Q^b \over \partial q^n}\,,
\eeq
and  
\beq\label{detOmega}
\Om ( \bQ) =\det\Om_{ab} = g \left(det\left({\pa\bq\over\pa\bQ}
\right)\right)^2\,.
\eeq
Thus,
\beq\label{Omegameasure}
\sqrt{\Om} = \sqrt{g} J\,,
\eeq
where 
\beq\label{jacobian}
J = det\left({\pa\bq\over\pa\bQ}
\right)
\eeq 
is the Jacobian of the transformation, rendering the measure invariant
\beq\label{invariantmeasure}
\sqrt{\Om}\,d\bQ = \sqrt{g}\,d\bq\,.
\eeq 
Finally, the hamiltonian in the new 
coordinates is written 
\beq\label{hQ}
H_Q = -{\hbar^2\over 2} \nabla_Q^2 + \tilde V(\bQ)\,,
\eeq
where 
\beq\label{Qlaplacian}
\nabla_Q^2 = {1\over\sqrt{\Om}}{\pa\over\pa Q^a}\left(\Om^{ab}\sqrt{\Om}
{\pa\over\pa Q^b}\right)
\eeq
is the invariant laplacian in the new coordinates $Q$, and $\tilde V(\bQ) = 
V(\bq(\bQ))$ (i.e., the potential is a scalar function under 
(\ref{qQtransformation})).

Wave functions also transform as scalars,
$\psi(\bq)\rightarrow \tilde\psi(\bQ) = \psi (\bq(\bQ))$, since amplitudes 
must remain invariant against the coordinate transformations: 
\beq\label{invariance}
\langle \psi_1|\psi_2\rangle = \int\limits_{\cal C} 
\sqrt{\Om}\,d\bQ \,\psi_1^*(\bq(\bQ))\,\psi_2(\bq(\bQ)) = 
\int\limits_{\cal C} 
\sqrt{g}\,d\bq \,\psi_1^*(\bq)\,\psi_2(\bq)\,.
\eeq
The form (\ref{hQ}) of the transformed hamiltonian, and the fact that 
wave functions are scalars, guarantee the invariance of matrix elements 
$\langle \psi_1|H|\psi_2\rangle $ under coordinate transformations. 

Let us assume that for one reason or another, the $Q$-coordinates represent 
the physical picture more transparently, and thus, working with $H_Q$ has 
some extra merit. $H_Q$ 
is symmetric with respect to the measure $\sqrt{\Om}d\bQ$. Thus, part
of the price to be paid working with the $Q$-coordinates is the need to drag 
along that pesky measure everywhere. Life would be simpler if we could rid 
ourselves from that measure and map $H_Q$ onto an effective hamiltonian 
$H_{eff}$ which is symmetric with respect to the flat measure $d\bQ$. 
This we can achieve by performing a similarity transformation. 
Observing that 
$$ \langle \psi_1|H|\psi_2\rangle  = 
\int\limits_{\cal C} 
\sqrt{\Om}\,d\bQ \,\psi_1^*(\bQ)\, H_Q \psi_2(\bQ) = 
\int\limits_{\cal C} \,d\bQ
\left(\Om^{1\over 4}\,\psi_1^*(\bQ)\right)\, 
\left(\Om^{1\over 4}\,H_Q\, \Om^{-{1\over 4}}\right)\,\left(\Om^{1\over 4}\,
\psi_2(\bQ)\right)
$$
it is obvious that required transformation is 
\beq\label{Heff}
H_{eff} = \Om^{1\over 4}\,H_Q\, \Om^{-{1\over 4}}
\eeq
with the appropriate transformation
\beq\label{wavefuntransformation}
\tilde\psi(\bQ) = \Om^{1\over 4}\,\psi(\bQ)
\eeq
of wave functions. Let us now massage $H_{eff}$ into a more transparent form.
It is a matter of straightforward calculation to show that 
\beqra\label{auxiliary}
\Om^{1\over 4}\,\nabla_Q^2\, \Om^{-{1\over 4}}  &=& 
\Om^{-{1\over 4}}\,{\pa\over\pa Q^a}\left(\Om^{ab}\sqrt{\Om}
{\pa\over\pa Q^b}\right)\, \Om^{-{1\over 4}}\nonumber\\
&=&\left(\Om^{-{1\over 4}}\,{\pa\over\pa Q^a} \Om^{1\over 4}\right)\Om^{ab}
\left(\Om^{1\over 4}{\pa\over\pa Q^b}\Om^{-{1\over 4}}\right)\nonumber\\
&=&{\pa\over\pa Q^a}\Om^{ab}{\pa\over\pa Q^b} - \left({\pa\over\pa Q^a}(
\Om^{ab}\,C_b)\right) - C_a\,\Om^{ab}C_b
\eeqra
where we have defined 
\beq\label{C}
C_a = {1\over 4} (\log \Om)_{,\,a}
\eeq
and where $(\cdot)_{,\,a}$ indicates a derivative with respect to $Q^a$. 
The operator $\Om^{1\over 4}\,\nabla_Q^2\, \Om^{-{1\over 4}}$ is manifestly 
symmetric with respect to the flat measure  $d\bQ$, as is evident in each
of the lines in (\ref{auxiliary}). Thus, $H_{eff}$ is indeed the desired 
hamiltonian we set out to find, which, following (\ref{auxiliary}), we may 
write explicitly as
\beq\label{Hfinal}
H_{eff} = {1\over 2} P_a\,\Om^{ab}\,P_b + {\hbar^2\over 2} C_a\,\Om^{ab}\,C_b
+  \tilde V(\bQ) + {\hbar^2\over 2}\left(\Om^{ab}\,C_b\right)_{,\,a}
\eeq
where we introduced the momentum operators 
\beq\label{momentumops}
P_a = -i\hbar {\pa\over\pa Q^a}\,.
\eeq
The terms in (\ref{Hfinal}) quadratic in $\hbar $ may be thought of as
a generalization of the centrifugal barrier which arises in the radial 
hamiltonian in $D$ dimensions\footnote{The radial part of the $D$-dimensional 
laplacian $\nabla_r^2 = r^{-(D-1)}\pa_r (r^{D-1}\pa_r)$, defined with
respect to the measure $r^{D-1}dr$, may be transformed by a similarity 
transformation into $\tilde\nabla_r^2 = r^{D-1\over 2} \nabla_r^2 
r^{-{D-1\over 2}} = \pa_r^2 - {(D-1)(D-3)\over 4r^2} = 
\pa_r^2 - C_r\Om^{rr}C_r$, which is defined with
respect to the flat measure $dr$.}. Evidently, these terms are purely a 
quantum mechanical effect.\footnote{As a side remark, we also mention 
that the last term in (\ref{Hfinal}) may be written as 
${\hbar^2\over 2}\left(\Om^{ab}\,C_b\right)_{,\,a} = 
-{\hbar^2\over 4}\,\Om^{1\over 2}\left(\nabla_Q^2\,\Om^{-{1\over 2}}\right)\,
.$} 

It is easy to see that
\beq\label{christoffel}
C_a = {1\over 2}\Gamma^b_{ba}
\eeq
where $\Gamma^a_{bc}$ is the second Christoffel symbol (i.e., the connection) 
of $\Om_{ab}$. However, sometimes a direct computation of the $C_a$ from their
definition (\ref{C}), or from the identity (\ref{christoffel}), may be too 
difficult to carry in practice. Thus, in order to bypass these potential 
difficulties, we shall now derive an identity satisfied by the $C_a$, from 
which we could compute them with somewhat less effort.

To this end we argue as follows: The invariant laplacian acting on a 
function which is a scalar under coordinate transformation produces yet 
another scalar function. Thus, 
\beq\label{laplaceianinvariance}
\nabla_q^2 \psi(q) = \nabla_Q^2 \psi(q(Q))\,.
\eeq
In particular, the coordinate functions themselves are scalars (their
differentials are one-forms). Let us define the quantities
\beq\label{omega}
\om^a =  - \hbar\nabla_q^2 Q^a = -\hbar\nabla_Q^2 Q^a\,.
\eeq
It follows from the definitions (\ref{omega}) and (\ref{C}) that 
$$\om^a = -\hbar\nabla_Q^2 Q^a = - {\hbar\over\sqrt{\Om}}
{\pa\over\pa Q^b}\left(\Om^{ab}\sqrt{\Om}\right) = -{\hbar\over 2}\Om^{ab}\,
(\log \Om)_{,\,b} - \hbar\Om^{ab}_{~,\,b}$$
or 
\beq\label{laplacianQ}
\om^a + 2\hbar\,\Om^{ab}C_b + \hbar\,\Om^{ab}_{~,\,b} = 0\,,
\eeq
which is the desired identity to determine the $C_a$. To simplify the 
computation, we are free to choose in (\ref{omega})
the coordinates $q^a$ in which the computation of 
$\om^a =  - \hbar\nabla_q^2 Q^a$ is as simple as possible.

This concludes our brief review and exposition of point canonical 
coordinate transformations in quantum mechanics, which we will use in 
the second part of this section to construct the collective 
hamiltonian of our normal matrix model.

The discussion of point canonical transformations in section 6.1 
in \cite{sakbook} was presented\footnote{We should mention that there is 
a typographical error in Eq.(6.19) in \cite{sakbook}. Its correct form 
is the sum of anticommutators $H_{eff} - H_{eff}^\dgg = {i\over 2}
\{\om^a + 2\hbar\,\Om^{ab}C_b + \hbar\,\Om^{ab}_{~,\,b}, P_a\}$. Also note, 
that the last (total divergence) term in our (\ref{Hfinal}) is missing 
from the analogous equation, Eq.(6.21), in \cite{sakbook}.} for the special 
case of flat euclidean metric $g_{ab} = \delta_{ab}$. For this metric, it is 
a straightforward exercise to show that $\om^a$, $C_a$ and $\Om^{ab}$ which we 
defined here for an arbitrary metric, coincide with their counterparts 
in chapter 6 of \cite{sakbook}. In addition, in that metric, our expression 
(\ref{Hfinal}) for $H_{eff}$ and the identity (\ref{laplacianQ}) reduce to 
their counterparts, Eqs.(6.21) and (6.20), respectively, in \cite{sakbook}.

\subsection{The Collective Field Hamiltonian for Normal Matrices}

The ground state, as well as other $U(N)$-singlet eigenstates of the 
hamiltonian (\ref{hamiltonian}), are totally symmetric functions 
$\chi_s(z_1,z_1^*,\ldots,z_n,z_n^*)$ of the eigenvalues. As we saw in the 
introduction, these eigenvalues comprise a two-dimensional system of 
interacting bosons.

Let  
\beq\label{ri}
\br_i = x_i\hat \bx + y_i\hat \by
\eeq
be the position operator of the $i$th eigenvalue $z_i=x_i + iy_i$ in the 
plane. The density operators
\beq\label{density}
\phi (\br)  = {1\over N} \sum_{i=1}^N\,\delta\,(\br - \br_i)\,,
\eeq
form a continuous basis (parametrized by $\br$) of commuting operators which 
are symmetric in the operators $\br_i$. Clearly, any operator
which is purely a symmetric combination of $\br_i$'s may be expressed in terms
of the density operators (\ref{density}). Thus, it is 
conceivable that any singlet wave function $\chi_s$ in our model could be 
expressed as functional of the eigenvalues of these density operators, 
namely, as a functional of the density function, which we also denote by 
$\phi(\br)$. The density function is evidently positive, 
\beq\label{positivity}
\phi(\br) \geq 0\,.
\eeq

This motivates us to apply the method of coordinate transformation which we
derived in the previous subsection, and make a transformation from the 
eigenvalues $z_i$ (or the corresponding commuting position operators $\br_i$),
which live in a configuration space with the non-euclidean metric 
(\ref{linelement}), to the set of commuting density operators. 
The latter are the {\em collective variables} for our system of bosons - they 
are the combinations of the 
original dynamical degrees of freedom, the $\br_i$, which are invariant under 
the permutation symmetry of the problem. The transformation we will perform 
affects only the singlet degrees of freedom, and leaves the $U(N)$-angular 
variables $R_{ij}$ unchanged, of course. 

In this section we will make the transformation to these collective variables 
and study the system of bosons in the large-$N$ limit. This method of 
transformation, which is commonly  known as the collective field method,  
was introduced in \cite{jevsak} (for a review, see chapter 7 of 
\cite{sakbook}). It should be thought of as a natural extension of the 
Bohm-Pines theory of high density plasma oscillations \cite{bohmpines}. 
An important application of this method in \cite{jevsak} was to study the 
singlet sector of the quantum hermitean matrix model. As was mentioned earlier, 
the same problem was solved earlier in \cite{bipz}, by mapping it onto a one
dimensional gas of noninteracting fermions. Here, we will extend these
works to normal matrices.

The number of degrees of freedom in the singlet sector of (\ref{hamiltonian}) 
is $2N$. In contrast, there is a continuum of density operators $\phi(\br)$, 
which are not all independent. For example, it is obvious that the constraint 
\beq\label{normalization}
\int\,d\br\,\phi(\br) = 1
\eeq
should hold. As our independent collective variables we choose the set 
of $2N$ Fourier modes
\beq\label{fourier}
\phi_\bk = \int\,d\br\,e^{-i\bk\cdot\br}\,\phi(\br) = {1\over N}\sum_{i=1}^N
\,e^{-i\bk\cdot\br_i}\,
\eeq
cutoff by $k_{max}$, $|\bk|\leq k_{max}$, where $\bk$ is properly  
discretized, e.g., by putting the bosons in a large box of linear 
size $L$, and imposing appropriate boundary conditions. 
The details of this discretization are not important for our discussion of 
the large-$N$ limit. Next, we assume that the eigenvalues condense
into a lump of linear size $L_c$, where $L_c << L$. The average bulk density 
of particles in the system is thus
\beq\label{bulkdensity}
\rho\sim {N\over L^2_c}\,, 
\eeq
and the microscopic interparticle distance will be of the order 
$l\sim 1/\sqrt{\rho}$. Thus the maximal Fourier components relevant to our
problem should be of the order $k_{max} = 1/l\sim \sqrt{\rho}$. We shall take 
the limit $N\rightarrow\infty$ together with 
$L\rightarrow\infty$ such that $\rho >>1$, or more precisely, $L>>L_c>>l$. 
Thus, the high density limit makes $k_{max}\rightarrow\infty$, and letting 
$L\rightarrow\infty$ makes $\bk$ continuous. 

In this limit, which we shall assume throughout our analysis, we can consider 
the continuum of $\phi(\br)$s as independent variables, constrained by 
(\ref{normalization}) (i.e., by $\phi_{\bk=0}=1$).

To summarize, we wish to transform the normal matrix hamiltonian 
(\ref{hamiltonian}) from the coordinates $q^a = \br_i, R_{ij}$ 
and metric $g_{ab}$  given by (\ref{linelement}), to the coordinates 
$Q^a = \phi_\bk, R_{ij} $ and metric $\Om_{ab}$ given by (\ref{Qmetric}), and 
to construct the {\em effective} hamiltonian (\ref{Hfinal}), according to the 
formalism developed in the previous subsection. 

Since this transformation does not affect the $U(N)$-angular coordinates 
$R_{ij}$ and does not mix them with the eigenvalues, $\Om^{ab}$ will have the
same block-diagonal structure as $g^{ab}$, and of course, their pure 
$U(N)$-angular sectors will coincide. We need only compute the transform of
the pure eigenvalue sector $\Om^{\phi_\bk,\phi_{\bk'}}$. 

We start by computing the latter, which we denote more conveniently by
$\Om(\bk,\bk';[\phi])$, with the functional dependence on $\phi$ indicated 
explicitly. Since (\ref{linelement}) is equivalent to
\beq\label{linelement1}
ds^2 = \sum_i d\br_i^2 + 2\sum_{i<j} dR_{ij}^*\,dR_{ij} \,|\br_i - \br_j|^2\,,
\eeq
we obtain from (\ref{inverseQmetric}) and (\ref{fourier}) that 
\beq\label{Omega}
\Om(\bk,\bk';[\phi])= \sum_i {\pa\phi_\bk\over \pa\br_i}\cdot 
{\pa\phi_{\bk'}\over \pa\br_i} = - {\bk\cdot\bk'\over N}\,\phi_{\bk + \bk'}\,.
\eeq
The effective hamiltonian (\ref{Hfinal}) contains also the quatities $C_a$, 
which we will determine from the identity (\ref{laplacianQ}). Thus, we have to 
compute $\om^{\phi_\bk} = \om(\bk;[\phi])$. From the definition (\ref{omega})  
we obtain\footnote{From this point throughout the rest of the paper we set
$\hbar=1$.} 
\beq\label{omgak1}
\om(\bk;[\phi]) =  - \nabla^2 \phi_\bk = -\nabla_s^2\,\phi_\bk\,,
\eeq
where $\nabla^2$ was defined in (\ref{nablasquare}). Thus, from 
(\ref{singletpart}), we find\footnote{This expression for $\om(\bk;[\phi])$ 
is the two-dimensional analog of Eq.(7.55) in \cite{sakbook} and Eq.(5.6) in 
\cite{jevsak}, namely, the expression $\om(k;[\phi]) = 
Nk^2\int\limits_0^1\,d\alpha \phi_{k\alpha}\phi_{k(1-\alpha)} = 
k^2\phi_k + {2ik\over N}\sum_ie^{-ik\lambda_i}\sum_j'
{1\over\lambda_i - \lambda_j}$ corresponding to the hermitean matrix model 
with eigenvalues $\lambda_i$.} 
\beqra\label{omegak2}
\om(\bk;[\phi]) &=& - {1\over |\Delta|^2}\sum_i\left[
{\partial\over\partial x_i}
\left( |\Delta|^2 {\partial\over\partial x_i}\right) + 
{\partial\over\partial y_i}\left( |\Delta|^2 {\partial\over\partial y_i}
\right)\right]\phi_\bk\nonumber\\{}\nonumber &=& 
\bk^2\phi_\bk +{i\bk\over N}\cdot\sum_i\,e^{-i\bk\cdot\br_i}\,
{\partial \log\,|\Delta|^2\over\partial \br_i}\,.
\eeqra
From (\ref{vandermonde}) we obtain 
$$ {\partial \log\,|\Delta|^2\over\partial \br_i} = 2\sum_j\!' {\br_i-\br_j
\over (\br_i-\br_j)^2}\,.$$

The latter sum has a simple physical interpretation in terms of 
two-dimensional electrostatics. If we think of the eigenvalues located at 
the points $\br_k$ in the plane as positive point unit charges, then  
\beq\label{electrici}
\bE^{(i)} (\br) = \sum_{j\neq i} {\br-\br_j
\over (\br-\br_j)^2}
\eeq
is the electric field at $\br$ due to all the other charges, except the one
at $\br_i$. Thus, 
\beq\label{omegak3}
\om(\bk;[\phi]) = 
\bk^2\phi_\bk +{2i\bk\over N}\cdot\sum_i\,e^{-i\bk\cdot\br_i}\,
\bE^{(i)} (\br_i)\,.
\eeq
We would like to avoid the $i$-dependence of the vector field 
$\bE^{(i)} (\br)$ in (\ref{omegak3}) in order to simplify the summation.   
The aim is to replace it by the full electric field 
\beq\label{electric}
\bE (\br) = \sum_j {\br-\br_j
\over (\br-\br_j)^2}\,.
\eeq
However, since $\bE (\br)$ is singular at $\br_i$ we have to replace 
$\bE^{(i)} (\br)$ by a regulated form of $\bE (\br)$, which will tend in 
the large density limit to the macroscopic smoothed electric
field, which corresponds to the smoothed macroscopic density 
$\phi(\br)$, namely, 
\beq\label{electric1}
\bE (\br) = N\int\,d\br'\,{\br - \br'\over (\br - \br')^2}\,\phi(\br')\,,
\eeq
and satisfies Gauss' law 
\beq\label{gauss}
\nabla\cdot \bE (\br) = 2\pi N\phi(\br)\,.
\eeq
That (\ref{electric1}) satisfies (\ref{gauss}) is a trivial consequence of
the identity 
\beq\label{deltaidentity}
\nabla\cdot {\br\over \br^2} =  2\pi\delta (\br)\,.
\eeq

To identify the regulated form, note first that 
\beq\label{electricis}
\bE^{(i)}_s (\br) = \sum_{j\neq i} {\br-\br_j
\over (\br-\br_j)^2 + s^2}\,,
\eeq
with $s$ a small real parameter (which we shall assume is much smaller than
the mean interparticle distance $\sim 1/\sqrt{\rho}$), should have a 
negligible effect on $\bE^{(i)} (\br)$ in a typical configuration of the 
eigenvalues, in which it is most likely that $|\br_i-\br_j|\geq 1/\sqrt{\rho}$
for $j\neq i$, due to electrostatic repulsion. In $\bE^{(i)}_s (\br_i)$ we can 
obviously relax the restriction $j\neq i$ in the summation. Thus, it is clear
that 
\beq\label{electricss}
\bE_s (\br) = \sum_j {\br-\br_j
\over (\br-\br_j)^2 + s^2}
\eeq
is the desired regulated form of the electric field, with 
$\bE^{(i)}_s (\br_i) = \bE_s (\br_i)$ to be substituted in (\ref{omegak3}). 

The electric field $\bE(\br) = {\br\over r^2}$ of a point unit charge at the 
origin is regulated in this way to $\bE_s(\br) = {\br\over r^2 + s^2}$, such 
that $\nabla\cdot \bE(\br) = 2\pi\delta(\br)\rightarrow\nabla\cdot \bE_s(\br) 
= 2\pi\delta_s(\br),$ where $2\pi\delta_s(\br) = {2s^2\over (r^2 + s^2)^2}$.  
As $s$ tends to zero, $\delta_s(\br)$ tends to $\delta (\br)$. As another 
motivation for introducing the regularization (\ref{electricss}), note that 
\beq\label{greensfunction}
E_{sx} - iE_{sy} = \sum_j {z^*-z_j^*\over |z-z_j|^2 + s^2} = 
\tr {(z-M)^\dgg\over (z-M)^\dgg (z-M) +s^2}\,,
\eeq
where $M$ is a normal matrix with eigenvalues $z_i$, is the regulated form 
of the Green's function $G(z,z^*) = \tr {1\over z-M}$ one would obtain from 
the method of hermitization of \cite{hermitization}.

$\bE_s (\br)$ is the interpolating regulated electric field which should 
tend, in the large density limit to a smooth asymptotic 
distribution, which we can obtain by solving (\ref{gauss}) for $\bE(\br)$, 
with the corresponding limiting smooth distribution of $\phi(\br)$ as the 
source term. Thus, in the large density limit we can replace  
$\bE^{(i)} (\br_i)$  in (\ref{omegak3}) by 
$\bE_s (\br_i)$ and write 
$$\om(\bk;[\phi]) = 
\bk^2\phi_\bk +{2i\bk\over N}\cdot\sum_i\,e^{-i\bk\cdot\br_i}\,
\bE_s(\br_i)\,.
$$
Henceforth we will suppress the index $s$ in $\bE_s (\br)$, with the limit $s
\rightarrow 0$ understood as the last limit taken.

Since $\bE(\br)$ is of order $N$, as can be seen from 
(\ref{electric}) (or (\ref{electric1})),
we should neglect the subleading $\cO(N^0)$ term $\bk^2\phi_\bk$ in the last 
equation. Therefore, to leading order in ${1\over N}$, 
\beq\label{omegak4}
\om(\bk;[\phi]) = 
{2i\bk\over N}\cdot\sum_i\,e^{-i\bk\cdot\br_i}\,\bE (\br_i)\,,
\eeq
which is a quantity of order $N$. 

Thus, we have determined $\Om(\bk,\bk';[\phi])$ and $\om(\bk;[\phi])$. 
The last ingredient needed for determining  $C_{\phi_\bk} = C(\bk;[\phi])$ 
from (\ref{laplacianQ}) is the divergence $\Om^{ab}_{~,\,b}$. Happily enough, 
it vanishes
\beq\label{nulldivergence}
\sum_{\bk'}\,{\delta \Om(\bk,\bk';[\phi])\over \delta \phi_{\bk'}} = 
- \sum_{\bk'}\,{\bk\cdot\bk'\over N}\,\delta(\bk) = 0\,,
\eeq
as can be seen from (\ref{Omega}). Substituting (\ref{Omega}), (\ref{omegak4})
and (\ref{nulldivergence}) in (\ref{laplacianQ}), we obtain the equation 
\beq\label{CeqFourier1}
\om(\bk;[\phi]) + 2\sum_{\bk'}\,\Om(\bk,-\bk';[\phi])\, C(\bk';[\phi]) = 0
\eeq
for $C(\bk;[\phi])$. In the combined large density and large volume limits
discussed above, the $\bk$-sums tend to Fourier integrals. Thus, in the 
limit
\beq\label{CeqFourier1}
\om(\bk;[\phi]) + 2\int {d\bk'\over (2\pi)^2}\,\Om(\bk,-\bk';[\phi])\, 
C(\bk';[\phi]) = 0\,,
\eeq
which we will transform now to $\br$-space. 

To this end we need 
\beq\label{omegar}
\om(\br;[\phi]) = \int {d\bk\over (2\pi)^2}\,e^{i\bk\cdot\br}
\om(\bk;[\phi])
= 2\nabla\cdot\left(\phi(\br)\bE(\br)\right)\,,
\eeq
where we used 
$$\int {d\bk\over (2\pi)^2}\,{1\over N}\sum_i e^{i\bk(\br-\br_i)}\bE(\br_i)
 = {1\over N} \sum_i\delta (\br-\br_i) \bE(\br) = \phi(\br)\bE(\br)\,.$$ 
We also need 
\beq\label{Omegar}
\Om(\br,\br';[\phi]) = \int {d\bk d\bk'\over (2\pi)^4}\,
e^{i\bk\cdot\br+i\bk'\cdot\br'}\,\Om(\bk,\bk';[\phi]) = {1\over N}\nabla_\br
\cdot\nabla_{\br'}\left(\phi(\br)\delta(\br-\br')\right)\,.
\eeq
Using (\ref{omegar}) and (\ref{Omegar}) we Fourier transform 
(\ref{CeqFourier1}) to $\br$-space and obtain
\beqra\label{Ceqrspace}
&& \om(\br;[\phi]) - {2\over N}\nabla\cdot\left(\phi(\br)\nabla
C(\br;[\phi])\right) = \nonumber\\
&&{2\over N}\nabla\cdot\left[\phi(\br)\left(
N\bE(\br) - \nabla C(\br;[\phi])\,\right)\right] = 0\,.
\eeqra
The minimal solution of this equation is simply
\beq\label{Csolution}
\nabla C(\br;[\phi]) = N\bE(\br)\,. 
\eeq
Thus, from (\ref{electric1}), $C(\br;[\phi])) = NU(\br)$, where $U(\br)$ is 
(minus) the electrostatic potential 
\beq\label{coulomb}
U(\br) = {1\over 2} \int\,d\br' \,\log \,(\br - \br')^2\,\phi(\br')\,.
\eeq
$\nabla C(\br;[\phi])$ in (\ref{Csolution}) coincides with the leading 
large-$N$ behavior of the analogous quantity in \cite{2dcscollective}. 
Also, it is the two-dimensional analog of the corresponding quantity obtained 
in the collective field formulation of the hermitean matrix 
model \cite{jevsak}, 
$${\partial C(x;[\phi])\over \partial x} = N\int\!\!\!\!\!\!- 
\,{\phi(y)\,dy\over x-y}\,,$$ 
which reproduces the results of \cite{bipz} correctly\footnote{In this 
context, we should also mention that Eq.(\ref{omegar}) (which leads, 
together with (\ref{Omegar}), to (\ref{Csolution})), is essentially the 
two-diemnsional analog of of Eq.(7.58) in \cite{sakbook} and Eq.(5.9)
in \cite{jevsak} for $\om(x;[\phi])$ in the hermitean matrix model. In those
references, the one-dimensional analog of of (\ref{omegar}) was obtained
directly from the expression for $\om(k;[\phi])$, which contained the 
subleading $k^2\phi_k$ term. However, the latter term was effectively lost 
by taking the continuum limit of the Fourier sum into a Fourier integral.} . 

In two dimensions, however, we have the freedom of adding 
a piece ${\bu(\br)\over \phi(\br)}$ to the right hand side of 
(\ref{Csolution}), where $\bu$ is an arbitrary divergence-free vector
\beq
\nabla\cdot \bu = 0\,.
\eeq
We shall not pursue this possibility here. 

We now have all the ingredients needed for constructing the effective collective 
hamiltonian (\ref{Hfinal}). From (\ref{Omegar}) and (\ref{Csolution}) we obtain 
the collective potential $V_{coll} = {1\over 2} C_a\,\Om^{ab}\,C_b $ as 
\beqra\label{Vcoll}
V_{coll} &=& {1\over 2} \int\,d\br\, d\br'\, C(\br)\,\Om(\br,\br')\,C(\br') = 
{1\over 2N}\int d\br \,\phi(\br)\, (\nabla C(\br))^2 \nonumber{}\nonumber\\
&=& {N\over 2}\int d\br\, \phi(\br)\, \bE(\br)^2\,,
\eeqra
which is a quantity of order $N^3$. In the collective field formulation
of hermitean matrices, this term can be written equivalently as an integral over 
$\phi^3(x)$ \cite{jevsak, sakbook}.

Similarly to (\ref{Vcoll}), the kinetic term ${1\over 2} P_a\,\Om^{ab}\,P_b$ in our 
model is 
\beq\label{kinetic}
K = {1\over 2} \int\,d\br\, d\br'\, \Pi(\br)\,\Om(\br,\br')\,\Pi(\br') = 
{1\over 2N}\int d\br\, \nabla\Pi(\br)\cdot \phi(\br) \nabla \Pi(\br)\,,
\eeq
where $\Pi (\br)$ is the momentum operator conjugate to $\phi(\br)$. $\phi(\br)$ 
and $\Pi(\br)$ satisfy the equal-time canonical commutation relation 
\beq\label{ccr1}
[\phi(\br), \Pi(\br')] = i\left(\delta (\br-\br') - {1\over L^2}\right)\,,
\eeq
where the inverse area subtraction arises because the zero-mode $\phi_{\bk =0}$
is non-dynamical, due to the constraint (\ref{normalization}).

Next, we have the divergence term 
${1\over 2}\left(\Om^{ab}\,C_b\right)_{,\,a} = -{1\over 4} 
(\om^a + \Om^{ab}_{~,\,b})_{,\,a}$, where we used (\ref{laplacianQ}). 
Thus, from (\ref{omegar}) and (\ref{nulldivergence}) we find it as
\beq\label{Vdivergence}
V_{divergence} = -{1\over 4}\int d\br\,{\delta \om(\br;[\phi])\over 
\delta \phi(\br)} = - {1\over 2}\int d\br\,\nabla\cdot \left({\delta \over 
\delta \phi(\br)} \,\left(\phi(\br)\bE(\br)\right)\right)\,.
\eeq
It is evidently a singular boundary contribution. Assuming boundary 
conditions such that $\phi(\br)$ vanishes at spatial infinity faster than 
$r^{-2}$, we obtain 
\beq\label{Vdivergence1}
V_{divergence} = -{1\over 2}\,\delta (\br =0)\,\int d\br\,
\nabla\cdot \bE(\br) = -N\pi\,\delta (\br =0)\,,
\eeq
where we used (\ref{gauss}). This result for $V_{divergence}$ is merely 
a universal constant shift of energy, {\em independent} of the external 
potential $V(\br^2)$. It is proportional to $N$, whereas the leading 
behavior of energy eigenvalues is $\cO (N^2)$, as we will see momentarily. 
Thus, to leading order in ${1\over N}$, we shall neglect this term from now on.

We should mention that this result for $V_{divergence}$ agrees, to leading 
order in ${1\over N}$, with the one obtained in Eq. (11) in 
\cite{2dcscollective} for the quadratic potential $V(\br^2) = 
{m^2\over 2}\,\br^2$. It was shown in \cite{2dcscollective} that this term 
was canceled against the divergent zero-point energy of the collective 
field $\phi (\br)$. This cancellation should occur for any 
$V(\br^2)$, since our result (\ref{Vdivergence1}) is manifestly independent 
of the external potential.

Finally, the contribution of the external potential is 
\beq\label{externalpotential}
V_{ext} = N\int d\br \,\phi(\br)\,V(\br^2)\,.
\eeq

Gathering all terms together, we obtain the collective hamiltonian as
\beq\label{Hcollective1}
H_{coll} = 
{1\over 2N}\int d\br\, \nabla\Pi(\br)\cdot \phi(\br) \nabla \Pi(\br)\, + 
{N\over 2}\int d\br\,\phi(\br)\, \bE(\br)^2\, +  N\int d\br \,\phi(\br)\,
V(\br^2)\,.
\eeq

In order to display the large-$N$ behavior of this theory explicitly, we 
rescale the coordinates as $\br = \sqrt{N} \bx $. This is just the statement
that $\tr M^\dgg M \sim N\int \br^2 \phi(\br) d\br$ is typically $\cO (N)$,
assuming $\phi( \sqrt{N} \bx)$ is supported in a disk $|\bx|<\cO(N^0)$. 
Consequently, $\phi(\br) = N^{-1} \vphi (\bx)$ (since $\phi(\br)\,d\br$ is 
invariant), and $\bE(\br) = \sqrt{N}\ce (\bx)$ (from (\ref{electric})), where
\beq\label{rescaledE}
\ce (\bx) = \int\,d\bx'\,{\bx - \bx'\over (\bx - \bx')^2}\,
\vphi(\bx')
\eeq 
is a quantity of $\cO (N^0)$.

Also, from (\ref{potential}), 
$V(\br^2) = V(N\bx^2) = N \cv(\bx^2)$, where 
\beq\label{scaledpotential}
\cv(\bx^2) = \sum_p g_p (\bx^2)^p\,.
\eeq 
Finally, in order that the momentum dependent terms scale as the
rest, we must rescale $\Pi(\br) = N^2 \pi(\bx)$. 
The rescaled form of the hamiltonian (\ref{Hcollective1}) is thus
\beqra\label{Hcollective}
H_{coll} &=& N^2\left\{\int d\bx\left(
{1\over 2}\nabla\pi(\bx)\cdot \vphi(\bx) \nabla \pi(\bx)\, + 
{1\over 2}\vphi(\bx)\, \ce(\bx)^2\,
+ \vphi(\bx)\,\cv(\bx^2)\,\right)\right.\nonumber\\
&-& \left.\lambda\left(\int d\bx\,\vphi(\bx) -1\right)
\right\}\,,
\eeqra
where we have added a lagrange multiplier to impose the constraint 
(\ref{normalization}). As expected, the hamiltonian (\ref{Hcollective}) and 
its eigenvalues, the ground state energy in particular, scale as $N^2$. 

The commutation relation of the rescaled canonically conjugate fields is 
\beq\label{ccr}
[\vphi(\bx), \pi(\bx')] = {i\over N^2}\left(\delta (\bx-\bx') - {N\over L^2}
\right)\,,
\eeq
where we used (\ref{ccr1}). This commutation relation 
indicates that $\vphi(\bx)$, $\pi(\bx)$ and the dynamics of 
(\ref{Hcollective}) become classical in the large-$N$ limit.

\subsubsection{The Hamiltonian Equations of Motion: Fluid Dynamics  
Interpretation}

The Heisenberg equations of motion (as well as their classical analogs) 
resulting from (\ref{Hcollective}) and (\ref{ccr}) 
are easily found to be $$\dot\vphi(\bx,t) = -{1\over 2}\nabla\cdot\left
(\{\vphi(\bx,t)\,,\nabla\pi(\bx,t)\}\right)\,, $$ 
i.e., 
\beq\label{phieq}
\dot\vphi(\bx,t) = -\nabla\cdot\left(\vphi(\bx,t)\nabla\pi(\bx,t)\right)\,,
\eeq 
and 
\beq\label{pieq}
\dot\pi(\bx,t) = -{1\over 2} (\nabla\pi(\bx,t))^2 - 
{\delta\over \delta \vphi(\bx)}\int d\bx'\left[{1\over 2}\vphi(\bx')\, 
\ce(\bx')^2\,+ \vphi(\bx')\,(\cv(\bx'^2)- \lambda)\right] + {N\over L^2} 
{\cal A}
\,,
\eeq
where 
\beq\label{Aterm}
{\cal A}  = \int\,d\bx\,\left[{1\over 2} (\nabla\pi(\bx))^2 + 
\int\,d\by\,{\delta\over \delta \vphi(\by)}
\left({1\over 2}\vphi(\bx)\, 
\ce(\bx)^2\,+ \vphi(\bx)\,(\cv(\bx^2)- \lambda)\right)
\right]
\eeq
is a constant. By assumption\footnote{See the discussion following 
(\ref{bulkdensity}).}, the coefficient ${N\over L^2}$ in front of 
${\cal A}$ in (\ref{pieq}) tends to zero, and we shall neglect it.

Eq. (\ref{phieq}) just means that the constraint (\ref{normalization}) is 
preserved by the dynamics. 

Let us take the gradient of (\ref{pieq}) so as to make the equations of 
motion more symmetric. The $\pi$-dependent terms in the resulting equation 
can be lumped together into the combination $\pa_t\nabla\pi + (\pa_\mu\pi)
\nabla (\pa_\mu\pi) = \pa_t\nabla\pi + (\nabla\pi)\cdot \nabla (\nabla\pi)$. 
Thus, the gradient of (\ref{pieq}) may be written as 
\beq\label{pieq1}
\pa_t\nabla\pi + (\nabla\pi)\cdot \nabla (\nabla\pi) = - \nabla w\,
\eeq
where 
\beq\label{enthalpy}
w(\bx;[\vphi]) = {\delta\over \delta \vphi(\bx)}\int d\bx'\left({1\over 2}
\vphi(\bx')
\, \ce(\bx')^2\,+ \vphi(\bx')\,\cv(\bx'^2) - \lambda \vphi(\bx')\right)\,.
\eeq
Evidently, we should interpret the equations of motion as describing the 
isentropic flow of a two-dimensional eulerian fluid, with density 
$\vphi(\bx,t)$, an irrotational velocity field ${\bf v}(\bx,t) = \nabla 
\pi(\bx,t)$, and enthalpy density $w(\bx;[\vphi])$ given by (\ref{enthalpy}). 
Eq.(\ref{phieq}) is the continuity equation
\beq\label{continuity}
\dot\vphi(\bx,t)  + \nabla\cdot\left(\vphi(\bx,t){\bf v}(\bx,t)\right) =0\,,
\eeq 
and (\ref{pieq1}) is just Newton's law 
\beq\label{fluid} 
\pa_t {\bf v}(\bx,t)
+ \left({\bf v}(\bx,t)\cdot \nabla\right){\bf v}(\bx,t) = - \nabla w
\eeq
for isentropic flow, since in such flows the adiabatic variation of the 
enthalpy is related to the variation of the pressure $P$ according to 
$dw_{_S}  = VdP = dP/\vphi $. 
The fluid dynamical interpretation of the collective field 
equations of motion (\ref{phieq}) and (\ref{pieq}) is by no means special to 
our model. As we have already mentioned in the introduction, there is an 
analogous interpretation of the collective field of hermitean matrices in 
terms of a one-dimensional fluid \cite{matytsin, polchinski}. This is 
certainly a generic feature of field theoretic hamiltonians with a kinetic 
term of the form $\int d\bx\,\nabla\pi(\bx)\cdot \vphi(\bx) \nabla \pi(\bx)\,.$

\pagebreak

\section{The Ground State Energy and Distribution of Eigenvalues}
\setcounter{equation}{0}
As was discussed in the previous section, $\vphi(\bx,t)$, $\pi(\bx,t)$ and the 
dynamics governed by (\ref{Hcollective}) become classical in the large-$N$ 
limit. More precisely, on general theoretical grounds \cite{yaffe}, the 
quantum theory defined by (\ref{hamiltonian}) (and (\ref{Hcollective})) 
contains a special set of generalized coherent states, which may be used to 
construct a classical phase space and to derive a classical hamiltonian, whose
resulting classical dynamics is equivalent to the large-$N$ limit of the 
original quantum theory. The large-$N$ ground state is then found by 
minimizing this classical hamiltonian. 

The solution of the classical equations of motion derived from 
(\ref{Hcollective}) for $\vphi(\bx,t)$ and $\pi(\bx,t)$, which minimizes $H_{coll}$, 
should be interpreted as the expectation values of these fields in the ground state 
of (\ref{Hcollective}). Evidently, the ground state (or vacuum) expectation 
value of $\pi(\bx,t)$ should be null, and that of $\vphi(\bx,t)$ 
should be time independent. Thus, setting $\pi_0(\bx,t) = 
\langle 0|\pi(\bx,t)|0\rangle =0$ and $\langle 0|\vphi(\bx,t)|0 \rangle = 
\vphi_0(\bx)$ into the classical equations of 
motion, we see that (\ref{phieq}) is satisfied trivially, while (\ref{pieq}) 
leads to the equation 
\beq\label{extremum}
{\delta\over \delta \vphi(\bx)}\int d\bx'\left({1\over 2}\vphi(\bx')\, 
\ce(\bx')^2\,+ \vphi(\bx')\,\cv(\bx'^2)
- \lambda \vphi(\bx')\right)_{|{\vphi_0}}\,= 0\,.
\eeq
Eq. (\ref{extremum}) is just the statement that $\vphi_0(\bx)$ is the 
minimum of the effective potential 
\beq\label{effectivepotential}
V_{eff} = \int d\bx\left({1\over 2}\vphi(\bx)\, \ce(\bx)^2\,
+ \vphi(\bx)\,\cv(\bx^2)
- \lambda \vphi(\bx)\right)\,.
\eeq
The external potential $\cv(\bx^2)$ in our model is rotation-invariant. Thus, 
we would expect that the eigenvalue distribution in the ground 
state should be rotation invariant as well. Furthermore, in 
a physically sensible model, the external potential $\cv(\bx^2)$ must be 
confining, in order to balance the electric repulsion of the 
eigenvalues. The compromise between repulsion and confinement should produce
a rotation-invariant charge density which is non-vanishing only inside a disk
of finite radius $R$.

In other words, it is sufficient to look for the minimum of 
(\ref{effectivepotential}) in the class of normalized density functions 
$\vphi$ which depend only on $r=|\bx|$, and vanish outside a disk of some 
radius $R$, which is fixed by the normalization condition 
\beq\label{radialnormalization}
2\pi\,\int\limits_0^R \vphi(r)\,r\,dr = 1\,. 
\eeq
A radial charge density $\vphi (r)$ is the source for a radial electrostatic 
field $\ce(\bx) = {\cal E}(r)\hat\br$ for which Gauss' law reads
\beq\label{gausslaw}
\nabla\cdot\ce = {1\over r}{\pa\over\pa r}\left(r {\cal E}(r)\right) = 
2\pi\vphi(r)\,,
\eeq
and from which it follows that 
\beq\label{radialvariation1}
{1\over r}{\pa\over\pa r}\left(r{\delta{\cal E}(r)\over\delta\vphi(r')}
\right) = 2\pi\delta (r-r')\,.
\eeq
We stress that (\ref{radialvariation1}) holds only {\em inside} the support
of $\vphi(r)$, i.e., only in the region $r\leq R$. For $r>R$, the radial 
field 
\beq\label{externalelectric}
{\cal E}(r)={1\over r}
\eeq
identically, for {\em any} $\vphi (r)$, rendering the variational derivative 
null in that region.  

Upon integrating (\ref{radialvariation1}) we obtain 
\beq\label{radialvariation}
r{\delta{\cal E}(r)\over\delta\vphi(r')} = 2\pi r'\theta (r-r')\,\theta (R-r)
\,,
\eeq
where we used the fact that ${\cal E}(r)$ arises only due to charges 
{\em inside} the disk of radius $r$ to fix the integration constant.

Thus, evaluating the effective potential (\ref{effectivepotential})
on the class of radially dependent densities, and taking the variation 
with respect to $\vphi(r)$, we obtain, using (\ref{radialvariation}),
\beq\label{radialvariationV}
{\delta V_{eff}\over\delta \vphi(r)} = 2\pi r\left[{1\over 2}{\cal E}^2(r)
+\int\limits_r^R 2\pi\vphi(r') {\cal E}(r')\,dr' + \cv(r^2) -\lambda
\right]\,.
\eeq
Therefore, the desired minimum condition is 
\beq\label{minimum1}
{1\over 2}{\cal E}^2(r)
+\int\limits_r^R 2\pi\vphi(r') {\cal E}(r')\,dr' + \cv(r^2) -\lambda =0\,.
\eeq
We can achieve considerable simplification of (\ref{minimum1}) by applying
$\pa_r$ to both its sides and then use Gauss' law (\ref{gausslaw}). We thus 
obtain the simple and elegant formula 
\beq\label{electrics}
{\cal E}^2(r) = r{\pa\over \pa r} \cv(r^2)\,,\quad\quad r\leq R
\eeq
for the electric field inside the condensate of eigenvalues.

By substituting the left-hand side of (\ref{gausslaw}) for $2\pi\vphi(r')$ 
in (\ref{minimum1}), integrating, and using (\ref{electrics}), we obtain
an equation for the Lagrange multiplier $\lambda$:
\beq\label{lambdaeq}
{1\over 2}{\cal E}^2(R) + \cv(R^2) -\lambda =0\,.
\eeq

Note from (\ref{electrics}) that since ${\cal E}^2(r)\geq 0$, the eigenvalues 
can condense {\em only} in regions where $\cv(r^2)$ is increasing (inside the 
disk $r\leq R$). This is obviously the result of electric repulsion among 
eigenvalues: in a region in which $\cv(r^2)$ is decreasing, the external 
force pushes the eigenvalues away from the origin towrds larger radii, where 
they can keep farther apart from each other, until they reach a region in 
which $\cv(r^2)$ is increasing and the external force is pointed inwards, 
which eventually confines them. This effect is demonstrated most clearly 
in the model in which the eigenvalues are confined inside a rigid disk of 
radius $R$, namely, $\cv(r^2) = 0$ for $r<R$ and infinite elsewhere. 
Evidently, in this case the eigenvalues are all stuck to the wall at $r=R$. 

A generic external potential  $\cv(r^2)$ might have several local wells 
inside the disk $r\leq R$, some (or all) of which might be occupied by 
eigenvalues. The eigenvalues will reside, of course, only in regions inside 
these wells in which $\cv(r^2)$ is increasing. Let us assume that $\vphi_K(r)$ 
is an extremal eigenvalue density in which there are $K$ occupied wells. 
Consider the $k$th well ($k=1, \ldots, K$). 
The eigenvalues will occupy in this well an annular region 
$r_k^{min}\leq r\leq r_k^{max}$ (and of course, $r_K^{max} = R$).  Thus, the 
eigenvalues in such a case will condense into a system of $K$ concentric 
annuli, with voids in between the annuli. Inside the $k$th annulus, the 
electric field is given by (\ref{electrics}). In the void $r_k^{max}\leq r
\leq r_{k+1}^{min}$ between the $k$th and $k+1$st annuli, the electric field 
is obviously 
\beq\label{interannuli}
{\cal E}_k^{({\rm void})}(r) = {\nu(r_k^{max})\over r}\,,
\eeq
where $\nu(r_k^{max}) = 2\pi\int\limits_0^{r_k^{max}}\vphi(r)\,rdr$ is the
fraction of eigenvalues confined up to $r_k^{max}$. For $k=K$, 
(\ref{interannuli}) coincides with  (\ref{externalelectric}). 
It follows from (\ref{gausslaw}) that ${\cal E}(r)$ is continuous wherever
$\vphi(r)$ is continuous. Thus, assuming $\cv(r^2)$ has continuous derivatives,
the electric field  ${\cal E}(r)$ should be continuous throughout the plane. 
The $K$-annular extremal eigenvalue density $\vphi(r)$ depends upon the $2K$ 
parameters\footnote{Note that once the $2K$ 
parameters $r_k^{min,max}$ are set, all the $\nu_k$ are determined, since 
inside the filled annuli $\vphi(r)$ is given by (\ref{gausslaw}) and 
${\cal E}(r)$ by (\ref{electrics}).} $r_k^{min,max}$, and the continuity 
conditions on the electric field at the $2K$ boundaries of the annuli provide 
the exact number of equations to determine these parameters. (The first of 
these conditions is of course ${\cal E}(r_1^{min}) =0 $, and the last one 
${\cal E}(r_K^{max} = R) = 1/R $ is equivalent to the overall normalization 
condition (\ref{radialnormalization}).)

Given the external potential mentioned above, the extremum condition 
(\ref{minimum1}) might have several solutions $\vphi_K(r)$, depending 
on the number $K$ of occupied wells. In order to decide which of these extrema
is the actual ground state configuration, we should evaluate the effective 
potential (\ref{effectivepotential}) at each of these extremal solutions and 
pick the the one with minimal energy. By tuning the couplings in $\cv(r^2)$ we
might induce {\em quantum} phase transitions between these possible 
multi-annular extremal configurations.

In this paper we shall not pursue the generic case any further, with its 
multi-annular eigenvalue density configuration, and the fascinating 
possibility of inducing quantum phase transitions in this kind of matrix
models. Instead, in what follows, we will solve for the ground state energy 
and eigenvalue distribution of the generic monotonically increasing external 
potentials, which necessarily yields a disk-like eigenvalue density 
configuration. However, in the explicit examples which will follow this 
exposition, we will compute the disk-annulus phase transition for the 
quartic potential.

Thus, assume  $\cv(r^2)$ is such that $\vphi_0(r)$ is voidless and supported 
in the disk $r\leq R$. The eigenvalues behave like positive point charges in 
our electrostatic desctiption of the matrix model. Thus, we should take the 
positive root of (\ref{electrics}) for 
\beq\label{electricfinal}
{\cal E}(r) =  \sqrt{r{\pa\over \pa r} \cv(r^2)}\,,\quad\quad r\leq R\,.
\eeq
From the continuity of ${\cal E}(r)$ at $r=R$ and (\ref{externalelectric}), 
we obtain 
\beq\label{Req}
R{\cal E}(R) =1\,,
\eeq
which together with (\ref{electricfinal}) forms an equation for $R$. 
(Equivalently, we can derive this equation by substituting Gauss' law 
(\ref{gausslaw}) into the normalization condition (\ref{radialnormalization})
.) Then, using (\ref{gausslaw}), we can determine the eigenvalue density in 
the ground state as 
\beq\label{phizero}
\vphi_0(r) =  {\theta (R-r)\over 2\pi r}{\pa\over\pa r}
\left(r^3 {\pa\over \pa r} \cv(r^2)\right)^{1\over 2}\,.
\eeq
From  (\ref{lambdaeq}) and (\ref{Req}) we obtain 
\beq\label{lambda1}
\lambda = {1\over 2R^2} + \cv(R^2)
\eeq
for the Lagrange multiplier $\lambda $. Finally, the ground state energy 
$E_0$ is found by substituting (\ref{electricfinal})-(\ref{phizero}) (and 
$\pi_0(\bx,t) = 0$) in the collective hamiltonian (\ref{Hcollective}), and 
using the normalization condition (\ref{radialnormalization}).  After some
algebra, we obtain  
\beqra\label{E0}
E_0 &=& N^2\int\limits_0^R\,dr \left({\pa\over\pa r}(r{\cal E}(r))\right)
\left({1\over 2}r{\pa \cv(r^2)\over \pa r} + \cv(r^2)\right)\nonumber\\
{}\nonumber\\
&=& N^2\int\limits_0^{R^2}\,du {\pa(u \cv(u))\over \pa u}\,
{\pa\over \pa u}\left(2u^2 {\pa\over \pa u} 
\cv(u)\right)^{1\over 2}\,.
\eeqra

\subsection{Examples}
We end this paper by applying the formalism developed in this section in 
two explicit examples. 
\subsubsection{The Quadratic Potential}
Consider the normal matrix model with quadratic potential (\ref{quadraticpotential1}),
namely, 
\beq\label{quadraticpotential}
\cv(r^2) = {1\over 2}m^2 r^2
\eeq
with $m^2>0$. This is essentially the model studied recently in detail in
\cite{2dcscollective}, and our purpose here is simply to show that the general
formalism developed above reproduces the results of 
\cite{2dcscollective} to leading order in ${1\over N}$.

From (\ref{electricfinal}) - (\ref{phizero}) we obtain
\beq\label{qudraticE}
{\cal E}(r) =  mr
\eeq
and 
\beq\label{qudraticphi}
\vphi_0(r) = {m\over \pi} = {\rm const.}
\eeq
in a disk of radius 
\beq\label{qudraticR}
R = {1\over \sqrt{m}}\,.
\eeq
Our result (\ref{qudraticphi}) is in agreement with Eq.(17) of 
\cite{2dcscollective} (with their $\lambda =1$). The ground state energy as 
found from (\ref{E0}) is
\beq\label{quadraticE0}
E_0 = {1\over 2} N^2 m\,,
\eeq
in agreement with (\ref{exactgsenergy}) and also with Eq.(15) of 
\cite{2dcscollective}, to leading order in ${1\over N}$.

\subsubsection{The Quartic Potential}
Next we consider the quartic potential, which gives rise to a nonuniform 
ground state eigenvalue distribution $\vphi_0(r)$. In this case the 
eigenvalues condense either into a disk or an annulus, depending on the sign
of the coefficient of the quadratic term. The disk phase occurs for a positive
coefficient of $r^2$, and the annular phase occurs for a negative 
coefficient.\\

\noindent{\bf (a)~The Disk Phase}\\
Consider the quartic potential 
\beq\label{quarticpotential}
\cv(r^2) = {1\over 2}m^2 r^2 + {g\over 4} r^4
\eeq
with $m^2,g>0$. From (\ref{electrics}), the electric field is  
\beq\label{quarticE}
{\cal E}(r) =  mr\left(1+{gr^2\over m^2}\right)^{1\over 2}
\eeq
and the density is
\beq\label{quarticphi}
\vphi_0(r) = {m\over \pi} {1+{3\over 2}{gr^2\over m^2}\over 
\sqrt{1+{gr^2\over m^2}}}\,,
\eeq
in a disk of radius $R$, where $R$ is obtained by solving (\ref{Req}). 
In the quartic case (\ref{Req}) leads to the cubic equation 
\beq\label{quarticR}
u^2(u+1) = a^2
\eeq
for the positive dimensionless quantities 
\beq\label{dimensionless}
u={gR^2\over m^2}\quad\quad {\rm and}\quad\quad a = {g\over m^3}\,.
\eeq
The physical root of (\ref{quarticR}) should be chosen so as to have to 
correct value in the weak coupling and strong coupling limits.

In the weak coupling limit $a = {g\over m^3}\rightarrow 0$, the 
electric field (\ref{quarticE}) and eigenvalue density (\ref{quarticphi}) tend 
manifestly to the corresponding quantities in the quadratic case. The root of 
(\ref{quarticR}) which tends smoothly to (\ref{qudraticR}) is $u\simeq a <<1$.

The strong coupling limit $a\rightarrow \infty$ also yields simple
expressions. The electric field tends to 
\beq\label{quarticEsc}
{\cal E}(r) =  \sqrt{g} r^2\,,
\eeq
the eigenvalue density tends to 
\beq\label{qudraticphisc}
\vphi_0(r) = {3\sqrt{g}\over 2\pi} r\,,
\eeq
and the radius of the disk tends to 
\beq\label{quarticRsc}
R= g^{-{1\over 6}}\,.
\eeq
The corresponding root of (\ref{quarticR}) is $u\simeq a^{2\over 3} >>1$. 

We shall not go into the solution of the cubic equation (\ref{quarticR}) for
arbitrary coupling in any more detail. However, it is worth mentioning that 
the weak coupling and strong coupling branches of the solution correspond to 
two different roots of (\ref{quarticR}) each of which becomes real in its turn, 
which cross at $a_* = \left({g\over m^3}\right)_* = {2\over\sqrt{27}}$.

Finally, the ground state energy as found from (\ref{E0}) is
\beq\label{quarticE0}
E_0 = {mN^2\over 4a^2}\left[{9\over 7}\left((u+1)^{7\over 2} -1 \right) 
-{9\over 5}\left((u+1)^{5\over 2} -1 \right) 
- {1\over 3}\left((u+1)^{3\over 2} -1 \right)+ 
\left((u+1)^{1\over 2} -1 \right)\right]\,.
\eeq
It is easy to check that in the weak coupling limit ($a, u<<1$) this 
expression tends to (\ref{quadraticE0}), as it should. In the strong coupling
limit ($a, u>> 1$), it tends to 
\beq\label{quarticE0sc}
E_0 = {9N^2 g^{1\over 3}\over 28} \left(1 + \cO(a^{-{2\over 3}})\right)\,,
\eeq
which can be verified by plugging $\cv(r^2) = {g\over 4} r^4$ directly into 
(\ref{E0}). The scaling of the ground state energy in the strong coupling 
limit as $g^{1\over 3}$ is expected, of course, due to simple dimensional 
analysis.\\

\noindent{\bf (b)~The Annular Phase}\\
In the annular phase the quartic potential is 
\beq\label{quarticpotentiala}
\cv(r^2) =- {1\over 2}\mu^2 r^2 + {g\over 4} r^4\,,
\eeq
with $\mu^2,g>0$. Then, from (\ref{electrics}), the electric field is  
\beq\label{quarticEa}
{\cal E}(r) =  \mu r\left({gr^2\over \mu^2} - 1\right)^{1\over 2}
\eeq
and the density is
\beq\label{quarticphia}
\vphi_0(r) = {\mu\over \pi} {{3\over 2}{gr^2\over \mu^2} -1\over 
\sqrt{{gr^2\over \mu^2} - 1}}\,,
\eeq
in an annulus $r_{min} \leq r \leq R$.

The radius of the inner boundary 
\beq\label{rmin}
r_{min}^2 = {\mu^2\over g}
\eeq
is found from the requirement that ${\cal E}^2(r)$ be positive. Note that near 
the inner boundary the density diverges as 
\beq\label{quarticphiasing}
\vphi_0(r) = {g^{-{1\over 4}}\over \pi} \left({\mu\over 2}\right)^{3\over 2}\,
{1\over \sqrt{r - r_{min}}} + \cO (\sqrt{r - r_{min}})\,.
\eeq
The outer radius $R$ is obtained by solving (\ref{Req}). The latter leads, in this 
case, to the cubic equation 
\beq\label{quarticRa}
u^2(u-1) = a^2
\eeq
for the positive dimensionless quantities 
\beq\label{dimensionless}
u={gR^2\over \mu^2}\quad\quad {\rm and}\quad\quad a = {g\over \mu^3}\,,
\eeq
in a similar manner to the disk phase.

As in the disk phase, the physical root of (\ref{quarticRa}) should be chosen so as to 
have the correct value in the weak coupling and strong coupling limits.

In the annular phase, unlike the disk phase, the weak coupling limit 
$a = {g\over \mu^3}\rightarrow 0$ is singular, since at $g=0$, the potential 
(\ref{quarticpotentiala}) is unbounded from below and monotonically decreasing.
The corresponding root of (\ref{quarticRa}) in this limit is 
$u = 1 + a^2 +\cO(a^4)$ which implies $R = r_{min} (1+ {a^2\over 2} + \cO(a^4))$. Thus, 
in the weak coupling limit the eigenvalues condense into a narrow annulus of very large radius 
$\simeq r_{min} = {1\over \sqrt{\mu a}}$ and width $r_{min}({a^2\over 2} + \cO({a^4})) \simeq
{1\over 2}\left({a^3\over \mu}\right)^{1\over 2}$.

The strong coupling limit $a\rightarrow \infty$ is essentially the same as in the
disk phase, as given by (\ref{quarticEsc}) - (\ref{quarticRsc}), and the 
corresponding root of (\ref{quarticRa}) is $u\simeq a^{2\over 3} >>1$. 
In particular, note that in the strong coupling limit, the radius $r_{min}$ of the 
inner boundary shrinks like $g^{-{1\over 6}}a^{-{1\over 3}} = Ra^{-{1\over 3}} $, and the 
coefficient of the singularity (\ref{quarticphiasing}) near the inner boundary diminishes like 
$g^{1\over 4}a^{-{1\over 2}}$, so that the voidless bounded density 
(\ref{qudraticphisc}) will be obtained in the limit.

We shall not go into the solution of the cubic equation (\ref{quarticR}) for
arbitrary coupling in any more detail, except for mentioning the fact that 
(\ref{quarticRa}) has only one real root, which interpolates smoothly between 
the weak and strong coupling limits.

Finally, the ground state energy in the annular phase can be computed in a
manner similar to the derivation of (\ref{E0}). One finds
\beq\label{E0a}
E_0  = N^2\int\limits_{r_{min}^2}^{R^2}\,du {\pa(u \cv(u))\over \pa u}\,
{\pa\over \pa u}\left(2u^2 {\pa\over \pa u} 
\cv(u)\right)^{1\over 2}\,,
\eeq
which leads to the explicit expression 
\beq\label{quarticE0a}
E_0 = {\mu N^2\over 4a^2}\left[{9\over 7}(u-1)^{7\over 2}  
+{9\over 5}(u-1)^{5\over 2} - {1\over 3}(u-1)^{3\over 2} - 
(u-1)^{1\over 2}\right]\,.
\eeq
In the strong coupling limit ($a, u>> 1$), it tends to (\ref{quarticE0sc}),
namely the strong coupling limit of the disk phase. 

In the weak coupling limit ($a<<1, u\simeq 1$), on the other hand, 
(\ref{quarticE0a}) tends to the singular limit 
\beq\label{quarticE0wca}
E_0 = -{\mu N^2 \over 4a} \left(1 + \cO(a^2)\right) \simeq 
-{N^2 \mu^4 \over 4g}\,.
\eeq
Thus, in the weak coupling limit, $E_0$ tends to $-\infty$, symptomatic 
of the fact that the model (\ref{quarticpotentiala}) does not exist at $g=0$. \\

\noindent{\bf (c)~A Few Comments on the Disk-Annulus Phase Transition}\\
Starting from a given point $(m^2,g)$ in the disk phase, with the potential 
(\ref{quarticpotential}), we can induce a {\em quantum} phase transition into the 
annular phase, by holding $g$ fixed and tuning $m^2$ down all the way through 
$m^2 =0$ into negative values. The electric field ${\cal E}(r)$, eigenvalue 
density $\vphi_0(r)$ and ground state energy $E_0$ are all continuous through the 
transition point at $m^2=\mu^2 =0$. At the transition, they are given by the strong 
coupling formulas (\ref{quarticEsc})-(\ref{quarticRsc}) and (\ref{quarticE0sc}). 
Moreover, the annular eigenvalue distribution starts with a vanishing inner radius 
$r_{min}$ at the transition.

This behavior is also expected of the disk-annulus phase transition in the
zero dimensional (i.e., time independent) quartic ensemble of normal matrices. 

It is also interesting to compare the disk-annulus phase transition here with the 
corresponding transition in the zero dimensional quartic ensemble of complex 
matrices, studied in \cite{fsz} and \cite{fz}. In the latter case, the Green's 
function (the resolvent) for the complex matrix M (i.e., the electric field, see 
(\ref{greensfunction})), the eigenvalue distribution in the complex plane, 
the Green's function for $M^\dgg M$ and the density of singular values of $M$ 
related to it, and thus the ``free energy'' of the Dyson gas of singular values, 
were all continuous through the transition. However, unlike the situation here, the 
disk phase there invaded well into the $\mu^2 > 0$ region in parameter space, and 
moreover, at the transition, the disk fragmented into an annulus of {\em finite} 
critical inner radius. (This occurred by having a central part of the disk, of 
finite size, progressively depleted, as the transition point was approached from the disk
phse.) Also, the density of eigenvalues at the boundaries of either the disk or the annulus 
was always finite, unlike (\ref{quarticphiasing}).

\pagebreak

{\bf Acknowledgments}~~~ 
I thank Boris Shapiro for a discussion concerning fluctuations around the 
ground state. I also thank Alexander Abanov, Ilya Gruzberg, Razvan Teodorescu,
Paul Wiegmann and Anton Zabrodin for discussions on some aspects of the 
connection between this work and the normal matrix models \cite{fingering}, and also 
for turning my attention to \cite{fs}. Thanks are due also to Stjepan Meljanac
for useful correspondence and for turning my attention to \cite{mel}.


\begin{thebibliography}{99}

\bibitem{fingering} O.Agam, E. Bettelheim, P. B. Wiegmann and A. Zabrodin, 
Phys. Rev. Lett. {\bf 88}, 236802 (2002);\\
R. Teodorescu, E. Bettelheim, O. Agam, A. Zabrodin and P. B. Wiegmann, 
hep-th/0401165; hep-th/0407017\\      
P. Wiegmann, in {\sl Statistical Field Theories}, A. Cappelli and G. 
Mussardo (eds.), p. 337 (2002) ( Kluwer Academic Publishers) 
(cond-mat/0204254);\\
A. Zabrodin, Ann. Henri Poincare {\bf 4}, S851 (2003) (cond-mat/0210331).

\bibitem{tau} I. K. Kostov, I. Krichever, M. Mineev-Weinstein, P. B. Wiegmann
and A. Zabrodin, in {\sl Random matrices and their applications}, 
MSRI publications {\bf 40}, 285 (2001) (Cambridge University Press) 
(hep-th/0005259).  

\bibitem{string} S. Yu. Alexandrov, V. A. Kazakov and I. K. Kostov, Nucl. 
Phys. {\bf B667}, 90 (2003). 

\bibitem{qhe} L.-L. Chau and Y. Yu, Phys. Lett. {\bf A167}, 452 (1992);
L.- L. Chau and O. Zaboronsky, Proceedings in Memory of Professor Wolfgang
Kroll, ed. J.P. Hsu et al. (World Scientific, Singapore, 1997.)

\bibitem{oas}  G. Oas,  Phys. Rev. E{\bf 55}, 205 (1997).

\bibitem{chauzab} L.-L. Chau and O. Zaboronsky, Commun. Math. Phys {\bf 196}, 
203 (1998).

\bibitem{wigzab} P. B. Wiegmann and A. Zabrodin, J. Phys. A: Math. Gen. {\bf 36} 
3411 (2003); hep-th/0309253.

\bibitem{fsz} J. Feinberg, R. Scalettar and A. Zee, J. Math. Phys. {\bf 42}, 
5718 (2001).

\bibitem{fz} J. Feinberg and A. Zee, Nucl. Phys. {\bf B501}, 643 (1997).

\bibitem{fs} M. V. Feigel'man and M. A. Skvortsov, Nucl. Phys. {\bf B506}, 665 
(1997).

\bibitem{calogero} F. Calogero, J. Math. Phys. {\bf 10}, 2191, 2197 (1969); 
{\em ibid.} {\bf 12}, 419 (1971).

\bibitem{sutherland} Bill Sutherland, J. Math. Phys. {\bf 12}, 246 
(1971).

\bibitem{sla} B. I. Simons, P. A. Lee and B. L. Altshuler, Nucl. Phys. {\bf B409}, 
487 (1993).

\bibitem{cm} F. Calogero and C. Marchioro, J. Math. Phys. {\bf 14}, 182 (1973). 

\bibitem{2dcs} A. Khare and K. Ray, Phys. Lett. {\bf A230}, 139 (1997).

\bibitem{ghosh} P. K. Ghosh, J. Phys. A: Math. Gen. {\bf 34}, 5583 (2001).

\bibitem{poly} A. P. Polychronakos, Phys. Lett. {\bf B408}, 117 (1997).

\bibitem{mel} S. Meljanac, M. Milekovi\'c and A. Samsarov, Phys. Lett. {\bf B594}, 241 
(2004).

\bibitem{jevsak} A. Jevicki and B. Sakita, Nucl. Phys. {\bf B165}, 511 (1980).

\bibitem{sakbook} B. Sakita, {\sl Quantum Theory of Many-Variable Systems
and Fields}, Chapters 6 and 7 (World Scientific, Singapore, 1985).

\bibitem{feinberg} J. Feinberg, in preparation. 

\bibitem{ginibre} J. Ginibre, J. Math. Phys. {\bf 6}, 440 (1965).

\bibitem{2dcscollective} V. Bardek and D. Jurman, hep-th/0403194.

\bibitem{bipz} E. Br\'ezin, C. Itzykson, G. Parisi and J.-B. Zuber, Commun. 
Math. Phys {\bf 59}, 35 (1978).

\bibitem{matytsin} A. Matytsin, Nucl. Phys. {\bf B411}, 805 (1994).

\bibitem{polchinski} J. Polchinski, Nucl. Phys. {\bf B362}, 125 (1991). 

\bibitem{bohmpines}  D. Bohm and D. Pines,  Phys. Rev. {\bf 92}, 609 (1953).

\bibitem{hermitization} J. Feinberg and A. Zee, Nucl. Phys. {\bf B504}, 579 
(1997).\\
See also R. A. Janik, M. A. Nowak, G. Papp and I. Zahed, 
Nucl. Phys. {\bf B501}, 603 (1997).

\bibitem{yaffe} L. G. Yaffe, Rev. Mod. Phys. {\bf 54}, 407 (1982).~~
For a brief summary of this coherent-state picture, see section 2 of
T. A. Dickens, U. J. Lindquister, W. R. Somsky and L. G. Yaffe, Nucl. 
Phys. {\bf B309}, 1 (1988).~~ The latter paper, as well as \cite{jevsak} and 
\cite{bipz}, are reprinted in\newline
E. Br\'ezin and S. R. Wadia, {\sl The Large N Expansion in Quantum Field 
Theory and Statistical Physics} (World Scientific, Singapore, 1993).


\end{thebibliography}
\end{document}